\documentclass[twocolumn]{aastex631}
\usepackage{newtxtext,newtxmath}
\usepackage[T1]{fontenc}
\usepackage{ae,aecompl}
\usepackage{graphicx}% Include figure files
\usepackage{dcolumn}% Align table columns on decimal point
\usepackage{bm}% bold math
\usepackage{color}
\usepackage{natbib}
\usepackage{ulem,soul}
\usepackage{multirow}
\usepackage{longtable}
\usepackage{xspace}
\usepackage{nicefrac}

\newcommand{\B}[2]{\ensuremath{[\text{#1}/\text{#2}]}\xspace}
\newcommand{\Msun}{\ensuremath{M_\odot}}

\newcommand{\ch }{\ensuremath{\mathrm{M_{Ch}}}\xspace}

\newcommand{\erg}{\ensuremath{\mathrm{erg}}}

\defcitealias{Thibodeaux2024}{T24}
\defcitealias{xing2023Natur}{X23}
\defcitealias{skuladottir2024ApJ}{S24}
\begin{document}
\title[Origin of $\alpha$-poor Very Metal-Poor Stars]{Origin of $\alpha$-poor Very Metal-poor Stars}
%\title{Exploring the origin of $\alpha$-poor Very Metal-Poor Stars: Core-Collapse Supernova vs Supernova 1a }

\correspondingauthor{S. K. Jeena}
\email{jeenaunni44@gmail.com}

\correspondingauthor{Projjwal Banerjee}
\email{projjwal.banerjee@gmail.com}

\author{S. K. Jeena}
\affiliation{Department of Physics\\
Indian Institute of Technology Palakkad, Kerala, India}

\author{Projjwal Banerjee}
\affiliation{Department of Physics\\
Indian Institute of Technology Palakkad, Kerala, India}
%\date{February 2024}

\begin{abstract}
Among very metal-poor (VMP) stars, $\alpha$-poor VMP ($\alpha$PVMP) stars that have sub-solar values of ${\rm [X/Fe]}$ for Mg and other $\alpha$ elements are rare and are thought to have been formed from gas polluted by Type 1a supernova (SN 1a). However, recent analyses indicate that pure core-collapse supernova (CCSN) ejecta can also be a likely source.  We perform a detailed analysis of 17 $\alpha$PVMP  stars by considering six different scenarios relevant to the early Galaxy. We consider a single pair-instability supernova (PISN) and a single CCSN. Additionally, we consider the combination of ejecta from a CCSN with ejecta from another CCSN, a PISN, a near-Chandrasekhar mass (near-${\rm M_{Ch}}$) SN 1a, and a sub-Chandrasekhar mass (sub-${\rm M_{Ch}}$) SN 1a. A clear signature can only be established for sub-${\rm M_{Ch}}$ SN 1a with a near-smoking-gun signature in SDSSJ0018-0939 and a reasonably clear signature in ET0381. The majority ($82\%$) of $\alpha$PVMP stars can be explained by pure CCSN ejecta and do not require any SN 1a contribution. However, the combination of CCSN and sub-${\rm M_{Ch}}$ SN 1a ejecta can also explain most ($76\%$) of $\alpha$PVMP stars. In contrast, the combination of ejecta from CCSN  with near-${\rm M_{Ch}}$ SN 1a and PISN can fit $41\%$ and $29\%$ of the stars, respectively. The single PISN scenario is strongly ruled out for all stars. Our results indicate that $\alpha$PVMP stars are equally compatible with pure CCSN ejecta and a combination of CCSN and SN 1a ejecta, with sub-${\rm M_{Ch}}$ SN 1a being roughly twice as frequent as near-${\rm M_{Ch}}$ SN 1a. 
\end{abstract}

\section{Introduction}
Very metal-poor (VMP) stars ($\B{Fe}{H}\leq-2$) with sub-solar values of $\B{X}{Fe}$ for $\alpha$ elements such as Mg, Si, and Ca are known as $\alpha$-poor VMP ($\alpha$PVMP) stars. They are considered to be chemically peculiar compared to most VMP stars that are found to be enhanced in $\alpha$ elements with super-solar $\B{X}{Fe}$. Among VMP stars,  $\alpha$PVMP  stars are quite rare and are usually associated with stars forming from gas polluted by Type 1a supernova (SN 1a) that naturally have $\alpha$-poor ejecta~\citep{ivans2003ApJ,Li2022ApJ}. However, recent studies by ~\citet{jeena_CCSN2024,Jeena_SN_1a_2024,jeena_LAMOST_revisit_2024} have found that the abundance pattern in  $\alpha$PVMP  stars can be fit well by 
ejecta from core-collapse supernova (CCSN), that do not undergo substantial fallback during the explosion.
In fact, in most cases, the quality of fit from pure CCSN ejecta is comparable to the fit from the mixing of ejecta between CCSN and SN 1a.
For example, a recent study by ~\citet{Jeena_SN_1a_2024} found that $\alpha$-poor metal-poor ($\alpha$PMP) and $\alpha$VMP stars namely COS171, BD+80245, HE0533-5340, and SMSSJ034249-284215, that were previously found to have strong SN 1a signatures~\citep{mcwilliam2018ApJ,reggiani2023AJ}, could be fit almost equally well by pure CCSN ejecta with no discernible difference in the matched abundance pattern. The exception to this was the $\alpha$PVMP star SDSSJ0018-0939 which was found to be unique in terms of being the only star where the fit to the observed abundance pattern from mixing of ejecta from CCSN and SN 1a was clearly better than the fit from pure CCSN ejecta. Similarly, a recent reanalysis of the  $\alpha$PVMP  star LAMOST J1010+2358 based on the updated observed abundance by \citet{Thibodeaux2024,skuladottir2024ApJ} (hereafter \citetalias{Thibodeaux2024} and \citetalias{skuladottir2024ApJ}), \citet{jeena_LAMOST_revisit_2024} found that pure CCSN ejecta can provide an excellent fit that is comparable to the fit provided by the combination of ejecta from CCSN and SN 1a. 

In order to get a complete picture of the origin of $\alpha$PVMP stars, we consider all known  $\alpha$PVMP  stars from the  
SAGA database~\citep{SAGA} and recent literature that satisfy the criteria $\B{Mg}{Fe}+\sigma(\B{Mg}{Fe})<0$, where $\sigma(\B{Mg}{Fe})$ is the observation uncertainty. We identify 17 such stars and compare the observed abundance pattern of each  $\alpha$PVMP  star with various theoretical abundance patterns resulting from the ejecta from all possible sources that were operating in the early Galaxy, i.e., CCSN, pair-instability supernova (PISN), near-Chandrasekhar mass (near-${\rm M_{Ch}}$) SN 1a, and sub-Chandrasekhar mass (sub-${\rm M_{Ch}}$) SN 1a, in order to find the most likely source.

\section{Methods}
We adopt exact methods developed in \citet{Jeena_SN_1a_2024} for our analysis that was recently also employed by \citet{jeena_LAMOST_revisit_2024}. 
Briefly, the method involves matching the observed abundance pattern from four distinct nucleosynthetic sources which are PISN, CCSN, near-\ch SN 1a, and sub-\ch SN 1a. We consider six different scenarios resulting from these sources as follows,

\begin{enumerate}
    \item Single PISN: ejecta from a single Pop III PISN.
    \item Single CCSN: ejecta from a single Pop III CCSN that undergoes mixing and fallback.
    \item 2CCSNe: the combination of ejecta from two single Pop III CCSN, one of which undergoes mixing and fallback.
    \item CCSN+near-\ch: the combination of ejecta from a single Pop III CCSN (with mixing and fallback) and a near-\ch SN 1a.
    \item CCSN+sub-\ch: the combination of ejecta from a single Pop III CCSN (with mixing and fallback) and a sub-\ch SN 1a.
    \item CCSN+PISN: the combination of ejecta from a single Pop III CCSN (with mixing and fallback) and a Pop III PISN.
\end{enumerate}
Pop III PISN yields are adapted from~\citet{hegerwoosley2002} that include 14 progenitors with He core masses of $65\hbox{--}130\,\Msun$ which correspond to initial  masses of $\sim140\hbox{--}260\,\Msun$. The near-\ch SN 1a yields are adapted from the 3D delayed detonation model N100\_Z0.01 by~\citet{2013seitenzahl} with an initial metallicity of $0.01\,Z_\odot$ and a central density of $2.9\times10^9{\rm g\,cm^{-3}}$. The sub-\ch SN 1a yields are adapted from the 3D double detonation models by~\citet{2021gronow} with an initial metallicity of $0.001\,Z_\odot$ from 11 models with CO core masses of $0.8\hbox{--}1.0\,\Msun$ and He shell masses of $0.02\hbox{--}0.1\,\Msun$. The CCSN models include yields from CCSN ejecta from 93 Pop III progenitors of mass ranging from $10\hbox{--}30\,\Msun$ using 1D hydrodynamic code \textsc{kepler}~\citep{weaver1978presupernova,rauscher2003hydrostatic} as discussed in ~\citet{Jeena_SN_1a_2024} and \citet{jeena_LAMOST_revisit_2024}. For each progenitor, we consider explosion energy $E_{\rm exp}$ of $1.2\times10^{51}\,\erg$ and $1.2\times10^{52}\,\erg$. In addition, we also consider $E_{\rm exp}$ of $0.3\times10^{51}\,\erg$ and $0.6\times10^{51}\,\erg$ for all progenitors of mass $<12\,\Msun$.
For each progenitor, we consider two different choices of initial mass cut $M_{\rm cut,ini}$ named $S_4$ and $Y_e$ models. In the former, the $M_{\rm cut,ini}$ is chosen to be at the location where the entropy per baryon exceeds $4k_{\rm B}$, whereas, in the latter, it is chosen to be at the edge of the Fe core where there is a jump in $Y_e$.
The ejecta for each CCSN model is calculated by mixing and fallback model as discussed in~\citet{tominaga2007, ishigaki2014} and more recently in~\citet{jeenaCEMP2023}. According to this model, all material above a final mass cut $M_{\rm cut,fin}$ is completely ejected, whereas a fraction $f_{\rm cut}$ between $M_{\rm cut,ini}$
and $M_{\rm cut,fin}$ is ejected. The amount of any isotope ejected by the CCSN is parameterized by $f_{\rm cut}$ and $M_{\rm cut,fin}$, where the amount of material that falls back onto the central remnant is $\Delta M_{\rm fb}=\Delta M_{\rm cut}(1-f_{\rm cut})$ where $\Delta M_{\rm cut}=(M_{\rm cut,fin}-M_{\rm cut,in})$.

Scenarios involving the mixing of ejecta from CCSN with another source S2 are parameterised by a single parameter $\alpha$ given by
\begin{equation}
    \alpha =\frac{ M{\rm_{dil,CCSN}}}{M{\rm_{dil,CCSN}}+M{\rm_{dil,S2}}},
    \label{eq:alpha}
\end{equation}
where $M{\rm_{dil,CCSN}}$ and $M{\rm_{dil,S2}}$ are the effective dilution masses from CCSN and source S2, respectively. In order to be consistent with the metal mixing of SN ejecta in the early Galaxy~\citep{chiaki2018, magg2020minimum}, we impose a minimum value of dilution mass of $10^4\,\Msun$ for SN 1a and CCSN models with $E_{\rm exp}\leq1.2\times10^{51}\,\erg$. For CCSN models with $E_{\rm exp}=1.2\times10^{52}\,\erg$ and all PISN models, we impose a minimum dilution mass of $10^5\,\Msun$.
The best-fit model from each scenario is found by using a $\chi^2$ prescription that involves parameters $M_{\rm cut,fin}$ and $ f_{\rm cut}$ for scenarios involving CCSNe and $\alpha$ for scenarios involving two sources as discussed in~\citet{heger2010nucleosynthesis, jeenaCEMP2023, Jeena_SN_1a_2024}. 
For scenarios involving CCSN and another source S2, where S2 is either PISN or SN 1a, we quantify the relative contribution from each source to an element ${\rm X_i}$ by calculating the fraction $\eta ({\rm X_i})$ of the total elemental yield $Y_{\rm X_i}$ where $\eta_{\rm CCSN}({\rm X}_i)+\eta_{\rm S2}({\rm X}_i)=1$. Similar to~\citet{Jeena_SN_1a_2024}, we treat Sc as an upper limit for scenarios that involve CCSN ejecta as they are dominantly produced by neutrino-processed proton-rich ejecta \citep{siverding2020ApJ,wang2024ApJ} which is not modelled in our 1D calculations. 

It is important to note that the $\chi^2$ from the best-fit single CCSN model is guaranteed to be greater than or equal to the $\chi^2$ from the best-fit model from CCSN+S2 scenarios, where S2 is either SN 1a or PISN. This is because all the single CCSN models with no contribution from S2 correspond to $\alpha \approx 0$, i.e., all the single CCSN models are a subset of all possible models from any CCSN+S2 scenario.
For this reason, the $\chi^2$ for the best-fit 2CCSNe model gives a fair evaluation of whether pure CCSN ejecta can fit the abundance pattern better for a star than the best-fit CCSN+SN 1a model. We do not consider pure  SN 1a ejecta separately as a near $100\%$ contribution from SN 1a is already included in the CCSN+SN 1a scenario corresponding to $\alpha\approx 1$. We also do not consider combining ejecta from a single PISN and SN 1a as both are much rarer compared to CCSN which makes their simultaneous occurrence extremely unlikely especially before the gas in the early Galaxy is polluted by CCSN.

\section{Key abundance features of PISN, CCSN, and SN 1a}
Below we briefly discuss the key features of the elemental abundance pattern of various sources adopted in this study which is discussed in detail in~\citet{Jeena_SN_1a_2024}.

The abundance pattern of a PISN model is very sensitive to the He core mass. 
With increasing He core mass, the value of $\B{X}{Fe}$ for even elements from C to Ca decreases dramatically. For example, $\B{X}{Fe}$ is highly super-solar with values of $\sim +2$ for C to Ca for the $70\,\Msun$ He core model. In sharp contrast, from He core models of $\gtrsim125\,\Msun$, $\B{X}{Fe}$ is sub-solar for C to Mg and roughly solar for Si to Ca. At the same time, the magnitude of the odd-even effect in terms of $\B{X}{Fe}$ for elements from Ne to Sc increases dramatically from  $\sim 1$ dex for the $65\,\Msun$ He core model to $\gtrsim2$ dex for the $130\,\Msun$ He core model.

The CCSN models have the most diverse elemental abundance patterns that depend on the mass of the progenitor, explosion energy, and the details of mixing and fallback. Even though the ejecta from CCSN is usually associated with super-solar values of $\B{X}{Fe}$ for $\alpha$ elements, many of the CCSN models with negligible fallback lead to sub-solar values. This is particularly relevant for $\alpha$PVMP stars as highlighted in \citet{jeena_CCSN2024,Jeena_SN_1a_2024}. The models that undergo merger of the O-burning shell with the O-Ne-Mg shell, lead to highly super-solar values of $\B{X}{Fe}$ for elements from Si to Ca along with sub-solar values of $\B{Mg}{Fe}$.

The abundance patterns from SN 1a differ from both PISN and CCSN models.
In contrast to CCSN models, the production of lighter elements up to Al is negligible in both near-\ch and sub-\ch SN 1a models. In the near-\ch SN 1a model, $\B{X}{Fe}$ for $\alpha$ elements is sub-solar for elements up to Ti. On the other hand, the abundance pattern from Si to Fe peak varies significantly among the various sub-\ch SN 1a models that depend on both  CO core mass ($M_{\rm CO}$) and He shell mass ($M_{\rm He}$). The value of $\B{X}{Fe}$ for $\alpha$ elements vary from super-solar for low $M_{\rm CO}\sim 0.8$ to sub-solar for $M_{\rm CO}\gtrsim 0.1$. 
The contribution from He shell burning to the final ejecta is a key feature in the sub-\ch SN 1a models from \citet{2021gronow} adopted in this study where the incomplete Si burning resulting from He detonation leads to super-solar values of $\B{X}{Fe}$ for Ti--Cr that is not seen in any other source. 

\section{Results and Discussion} \label{sec:ch7_best-fit_resulls}
Based on the results from the best-fit analysis for each star, we broadly divide them into five groups corresponding to the scenario that provides the overall best-fit i.e., lowest $\chi^2$, as follows;

\begin{itemize}
    \item Group A: The best-fit 2CCSNe model is the overall best-fit. 
    \item Group B: The best-fit CCSN+near-\ch model is the overall best-fit. 
    \item Group C: The best-fit CCSN+sub-\ch model is the overall best-fit.
    \item Group D: The best-fit CCSN+PISN model is the overall best-fit.
    \item Group E: The best-fit single PISN model is the overall best-fit.
\end{itemize}
Because the best-fit single CCSN model for any star is guaranteed to have a $\chi^2$ greater than or equal to the best-fit model from any scenario involving CCSN and another source, we do not have a group that corresponds to the lowest $\chi^2$ from the best-fit single CCSN model.

Out of the 17 stars, we find that 8 belong to Group A, 2 to Group B, and 7 to Group C. We find that no stars belong to either Group D or Group E. For each star, we classify the quality of fit from the best-fit model from each scenario as very good, good, acceptable, poor, and very poor. This is based on the detailed analysis that takes into account the value of $\chi^2$, the number of elements that can be matched within $1\sigma$ uncertainty, the number of outliers and their level of deviation beyond $1\sigma$.
For all stars,  the best-fit models are plotted using the following line style and colour combinations: single PISN (magenta dotted line),   single CCSN (black dashed line), 2CCSNe (cyan dashed line), CCSN+near-\ch (red dashed-dotted line), CCSN+sub-\ch (blue dashed-dotted line), and CCSN+PISN (green dashed line). 

\subsection{Best-fit Group A Stars}
Out of the 17 stars, 8 stars belong to Group A. Below, we discuss the detailed analysis of the best-fit models from all scenarios for each of the stars. The stars are discussed in decreasing order of number of elements detected up to Zn.  The best-fit abundance plots are presented in Fig.~\ref{fig:groupA_1} and Fig.~\ref{fig:groupA_2} with the corresponding information about the best-fit models and parameters listed in Table~\ref{tab:best_fit_GA_1} and Table~\ref{tab:best_fit_GA_2}, respectively.

\begin{figure*}
    \centering
    \includegraphics[width=\columnwidth]{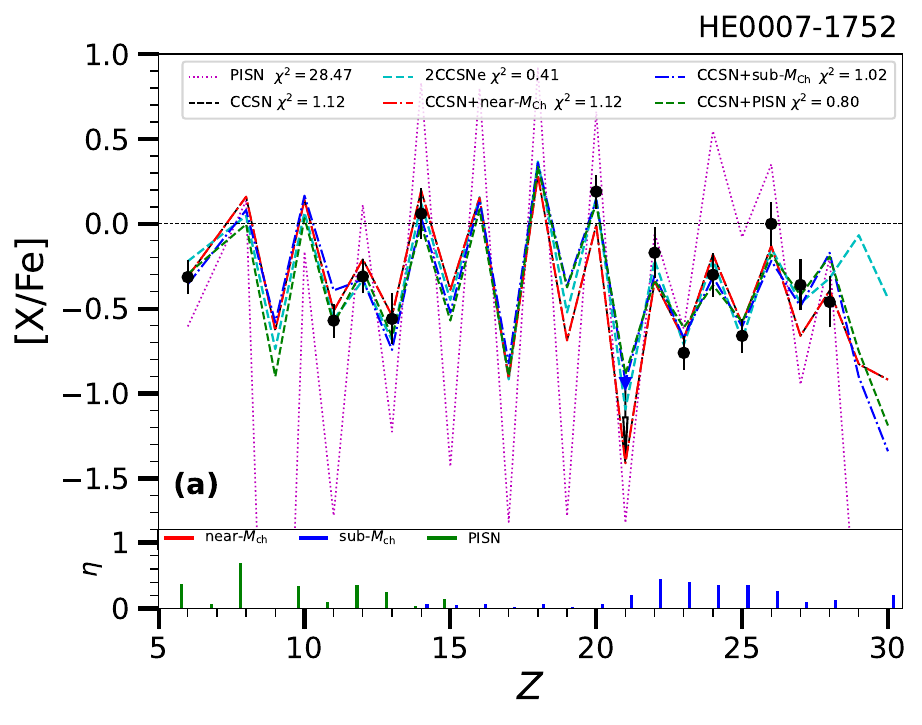}
    \includegraphics[width=\columnwidth]{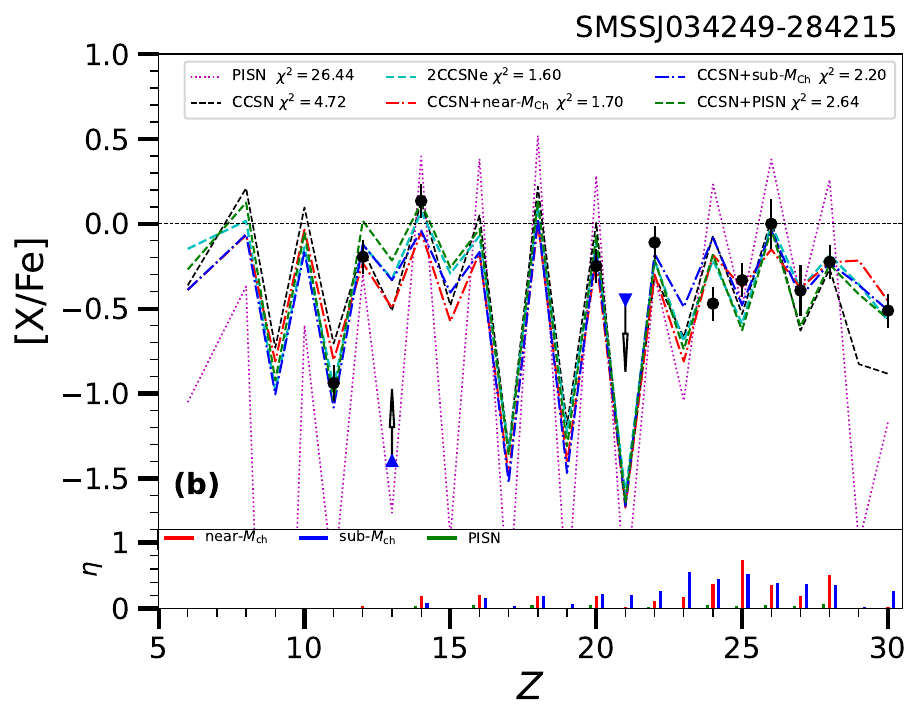}\\
    \includegraphics[width=\columnwidth]{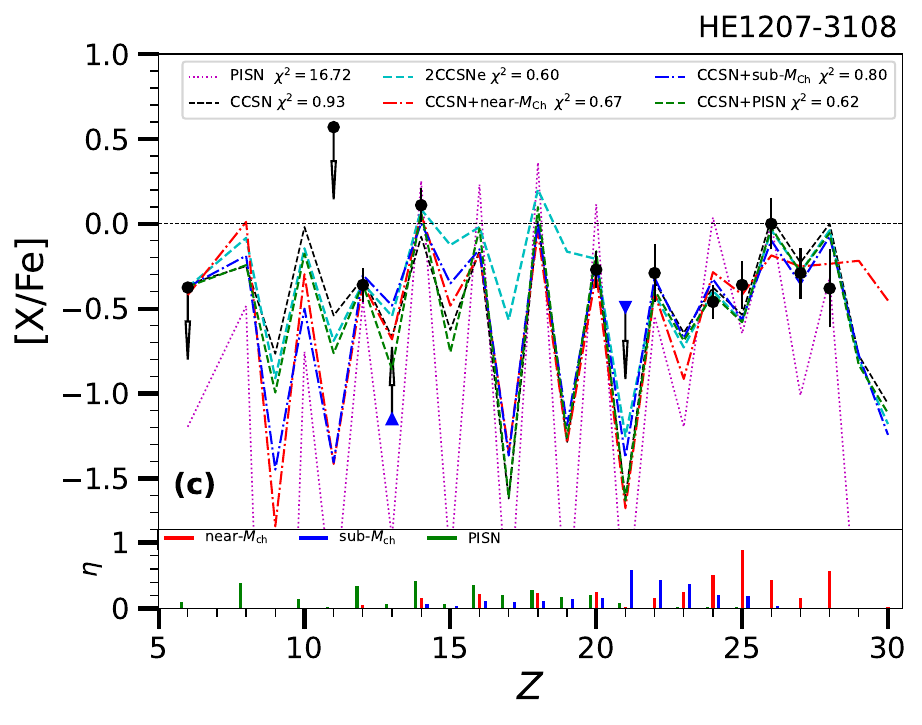}
    \includegraphics[width=\columnwidth]{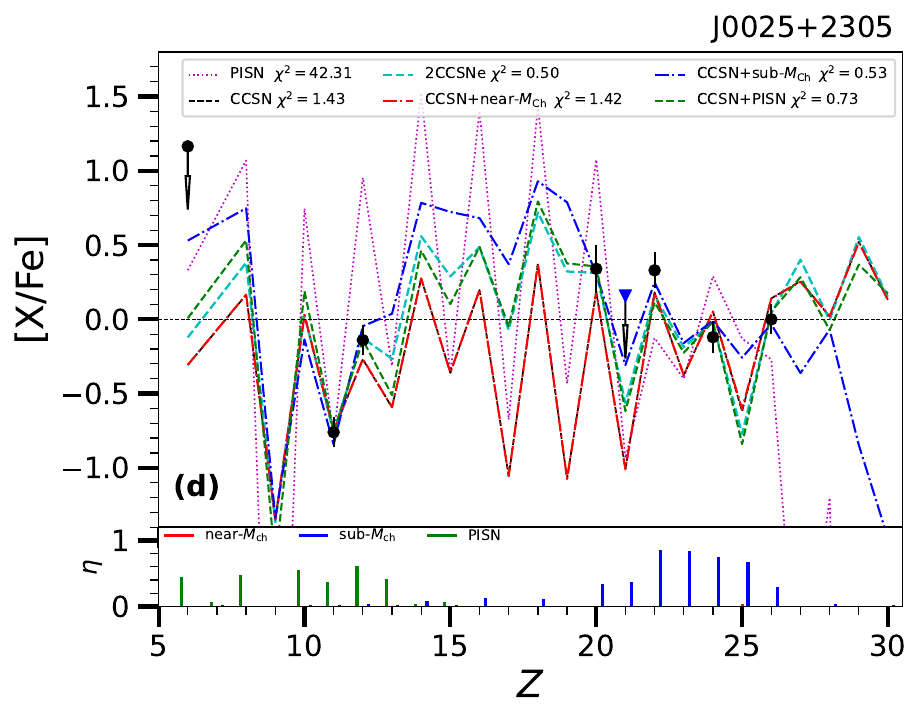}
    \caption{a) Top: The elemental abundance pattern of HE0007-1752~\citep{gull2021ApJ} compared with the best-fit models from various scenarios: single PISN (magenta dotted line), single CCSN (black dashed line), 2CCSNe (cyan dashed line),  CCSN+near-\ch  (red dashed-dotted line), CCSN+sub-\ch (blue dashed-dotted line), and CCSN+PISN (green dashed line). Bottom: The fraction $\eta$ for all elements produced by near-\ch SN 1a in CCSN+near-\ch scenario (red), sub-\ch SN 1a in CCSN+sub-\ch scenario (blue), and PISN in CCSN+PISN scenario (green). Note that Ti II abundance is adopted for Ti.
    b) Same as (a), but for SMSSJ034249-284215~\citep{reggiani2023AJ}. c) Same as (a), but for HE1207-3108~\citep{yong_norris2013}. d) Same as (a), but for J0025+2305~\citep{Li2022ApJ}. Note that the observed Sc is treated as an upper limit in all stars indicated using downward blue triangles. The observed Al is treated as a lower limit in SMSSJ034249-284215 and HE1207-3108 indicated using upward blue triangles.}
    \label{fig:groupA_1}
\end{figure*}

\subsubsection{HE0007-1752 (Fig.~\ref{fig:groupA_1}a)}

\begin{table*}
\centering
{\scriptsize\tabcolsep=3.0pt  % hold it local
\caption{Best-fit models and corresponding parameters along with the dilution mass from six scenarios compared to the observed data corresponding to the 4 Group A stars shown in Fig.~\ref{fig:groupA_1}. The outliers and the corresponding deviations from the central observed value (in units of $1\sigma$) are listed for all scenarios except for the single PISN scenario which has too many outliers.}
\label{tab:best_fit_GA_1}
\begin{tabular}{|c|c|c|c|c|c|c|c|c|r|}
\hline
    Star&Scenario&Model name&$E_{\rm exp}$ &$\chi^2$&$\alpha$&$\Delta M_{
    \rm cut}$&$\Delta M_{\rm fb}$&$M{\rm_{dil}}$&Outliers\\
    && & ($\times 10^{51}\,\erg$)& &&(\Msun) &(\Msun)& ($\times 10^4\,\Msun$)&\\
    \hline
    \multirow{9}{*}{\rotatebox[origin=c]{90}{HE0007-1752}}&PISN&$110\,\Msun$ He core&56.4 &28.47&--&--&--& $1.2\times10^4$&--\\
    \cline{2-10}
            &CCSN&\texttt{z19.8}-$S_4$&1.2&1.12&--&0.11 &0.00 &3.91&Mg (1.01) Ca (1.96), Ti (1.14), Co (2.00)\\
        \cline{2-10}
             &CCSN+&\texttt{z10.7}-$Y_e+$&1.2&\multirow{2}{*}{0.80} &\multirow{2}{*}{$4\times10^{-3}$} &0.10 &0.04 &1.07&\multirow{2}{*}{Ti (1.08), V (1.49), Fe (1.40), Ni (1.78)}\\ 
             &PISN&$65 \,\Msun$ He core & 4.9&&&-- & --&  $2.7\times10^2$&\\
        \cline{2-10}
            & \multirow{2}{*}{2CCSNe} & \texttt{z12.7}-$S_4$&1.2&\multirow{2}{*}{0.41}&\multirow{2}{*}{0.69}&0.00&0.00&2.37&\multirow{2}{*}{Fe (1.10)}\\
            && \texttt{z18}-$S_4$&1.2& & &0.19 &0.09 &5.81&\\
        \cline{2-10}
            & CCSN+&\texttt{z19.8}-$S_4 +$&1.2&\multirow{2}{*}{1.12}&\multirow{2}{*}{$10^{-7}$}&0.11 &0.00 &3.91&\multirow{2}{*}{Mg (1.01) Ca (1.96), Ti (1.14), Co (2.00)}\\
           &near-\ch&  N100\_Z0.01& --& && --&--&$3.9\times10^7$& \\
        \cline{2-10}
           &CCSN+&\texttt{z18}-$Y_e +$&1.2&\multirow{2}{*}{1.02}&\multirow{2}{*}{0.05}& 0.29&0.16 &3.69&\multirow{2}{*}{Na (1.78) Al (1.22), Fe (1.68), Ni (1.93)}\\
           &sub-\ch&   M10\_03&-- && &-- &-- &$7.0\times10^1$&\\
          % &&&&&&&&\\
            \hline
            \hline
    \multirow{9}{*}{\rotatebox[origin=c]{90}{SMSSJ034249-284216}}&PISN&$130\,\Msun$ He core&87.3 & 26.44&-- &--&-- & $1.5\times10^4$&--\\  
    \cline{2-10}
          & CCSN&\texttt{z23}-$S_{\rm 4} $&1.2&4.72& --& 0.67& 0.00&1.96&Na (2.21), Ca (2.56), Ti (1.97), Cr (3.99), Mn (2.00), Co (1.57), Zn (3.72)\\ 
        \cline{2-10}
         &CCSN+&\texttt{z16.4}-$S_4 $ +      &1.2 & \multirow{2}{*}{2.64}& \multirow{2}{*}{$2\times 10^{-4}$}& 0.29 &0.00  &1.01&\multirow{2}{*}{Mg (2.12), Ca ( 1.79), Ti (1.02), Cr (2.88), Mn (2.95), Co (1.42)} \\
         &   PISN                       &$130 \,\Msun$ He core            &87.3& &                                            &--    & --   & $5.1 \times 10^3$&\\
        \cline{2-10}
         &\multirow{2}{*}{2CCSNe}&\texttt{z13.7}-$S_{\rm 4} +$ &1.2& \multirow{2}{*}{ 1.6}& \multirow{2}{*}{ 0.63}& 0.00&0.00& 1.70&\multirow{2}{*}{Ca (1.36), Ti (1.04), Cr (2.57), Mn (2.57)}\\
         & &\texttt{z15.2}-$S_{\rm 4} $&1.2 & && 0.69&0.19 &1.00&\\ 
        \cline{2-10}
         &  CCSN+&\texttt{z15.2}-$S_{\rm 4} +$ &1.2& \multirow{2}{*}{1.7}& \multirow{2}{*}{0.05}&0.89 &0.00  & 1.07&\multirow{2}{*}{Na (1.29), Si (1.83), Ti (1.88), Cr (2.87), Fe (1.01)}\\
         &near-\ch&N100\_Z0.01&-- & & &--&--&$2.0\times10^1$ &\\ 
      \cline{2-10}
         &CCSN+ &\texttt{z14.8}-$S_4$&1.2& \multirow{2}{*}{2.20}& \multirow{2}{*}{0.06}& 0.20& 0.02 & 1.00 &\multirow{2}{*}{Na (1.38), Si (1.74), Cr (3.88), Mn (1.47)}\\
         &sub-\ch& M10\_10 &-- && &-- &--&$1.6\times10^1$ &\\ 
         \hline
         \hline
     \multirow{9}{*}{\rotatebox[origin=c]{90}{HE1207-3108}}&PISN&$120\,\Msun$ He core&71.0 &16.72&--&--&--& $1.1\times10^5$&--\\
     \cline{2-10}
    &CCSN&\texttt{z11.4}-$Y_e$&1.2&0.93&--& 0.19&0.06 &2.26&Si (1.86), Mn (1.26), Ni (1.66)\\
    \cline{2-10}
    &CCSN+&\texttt{z11.4}-$Y_e$&1.2&\multirow{2}{*}{0.62} &\multirow{2}{*}{0.009} & 0.49&0.29 &1.49& \multirow{2}{*}{Mn (1.57), Ni (1.41)}\\ 
    &PISN&$75\,\Msun$ He core +& 13.8&&&-- & --&  $1.6\times10^2$&\\
    \cline{2-10}
     & \multirow{2}{*}{2CCSNe} & \texttt{z11.4}-$Y_e +$&1.2&\multirow{2}{*}{0.60}&\multirow{2}{*}{0.07}&0.00&0.00&2.99&\multirow{2}{*}{Mn (1.54), Ni (1.49)}\\
    && \texttt{z19.6}-$Y_e$&1.2& & &0.19 &0.07 &$4.0\times10^1$&\\
    \cline{2-10}
    & CCSN+&\texttt{z19.8}-$Y_e+$&12&\multirow{2}{*}{0.67}&\multirow{2}{*}{0.09}& 1.06&0.87 &$1.0\times10^1$&\multirow{2}{*}{Cr (1.71), Fe (1.27)}\\
     &near-\ch&  N100\_Z0.01&-- & && --&--&$1.0\times10^2$ &\\
     \cline{2-10}
     &CCSN+&\texttt{z13.8}-$Y_e +$&1.2&\multirow{2}{*}{0.80}&\multirow{2}{*}{0.02}&0.29 &0.17 &4.80&\multirow{2}{*}{Si (1.06), Cr (1.34),  Mn (1.46), Ni (1.33)}\\
       &sub-\ch&   M08\_05& --&& &-- &--&$2.4\times10^2$ &\\
         \hline
         \hline
    \multirow{9}{*}{\rotatebox[origin=c]{90}{J0025+2305}}&PISN&$85\,\Msun$ He core&23.2 &42.31&--&--&--& $3.6\times10^3$&--\\
     \cline{2-10}
      &CCSN&\texttt{z19.2}-$Y_e$&12&1.43&--& 0.60& 0.13&$1.0\times10^1$&Mg (1.29), Ti (1.24), Cr (1.56), Fe (1.41)\\
       \cline{2-10}
     &CCSN+&\texttt{z21}-$Y_e+$&12&\multirow{2}{*}{0.73} &\multirow{2}{*}{0.03 }& 0.89&0.35 &9.90&\multirow{2}{*}{Ti (1.79)}\\ 
     &PISN&$65 \,\Msun$ He core & 4.9&&&-- & --&  $3.2\times10^2$&\\
      \cline{2-10}
     & \multirow{2}{*}{2CCSNe} & \texttt{z19.4}-$Y_e +$&12&\multirow{2}{*}{0.50}&\multirow{2}{*}{0.49}&0.00&0.00&$1.1\times10^1$&\multirow{2}{*}{Ti (1.41)}\\
           && \texttt{z17.8}-$Y_e$&12 & && 0.70&0.70 & $1.1\times10^1$&\\
    \cline{2-10}
       & CCSN+&\texttt{z19.2}-$Y_e+$&12&\multirow{2}{*}{1.42}&\multirow{2}{*}{$2\times10^{-3}$}& 0.60&0.13 &$1.0\times10^1$&\multirow{2}{*}{Mg (1.27), Ti (1.26), Cr (1.57), Fe (1.40)}\\
           &near-\ch&  N100\_Z0.01&-- & &&-- &--&$4.9\times10^3$ &\\   
     \cline{2-10}
      &CCSN+&\texttt{z19.6}-$Y_e +$&1.2&\multirow{2}{*}{0.53}&\multirow{2}{*}{0.03}&2.11 &1.96 &1.28&\multirow{2}{*}{Cr (1.01)}\\
    &sub-\ch&   M08\_05& --&& & --&--&$4.1\times10^2$ &\\  
     \hline
       \end{tabular}}%end
\end{table*}

HE0007-1752 has a metallicity of $\B{Fe}{H}=-2.36$ with 14 elements detected with $Z\leq30$~\citep{gull2021ApJ}. The abundance pattern from best-fit models and the observed abundances are shown in Fig.~\ref{fig:groupA_1}a. The overall best-fit 2CCSNe model provides a very good fit with $\chi^2=0.41$ and can match the abundances of all the elements within the observed $1\sigma$ uncertainty except Fe which is a minor outlier with a deviation of $1.10\sigma$. The best-fit model is comprised of a combination of ejecta from the \texttt{z12.7}-$S_4$ model without fallback and the \texttt{z18}-$S_4$ model with minimal fallback with standard explosion energy of $1.2\times 10^{51}\,\erg$. 
Among the remaining scenarios, only the single PISN scenario provides an extremely poor fit and can be ruled out.
The quality of fits from the rest of the scenarios can be classified as good but is slightly inferior compared to the 2CCSNe scenario as they have some additional outliers beyond the $1\sigma$ uncertainty but can match all elements within $2\sigma$ uncertainty. 
The contribution from sub-\ch SN 1a in the best-fit model is only significant for Ti--Fe with $\eta_{\rm 1a}\sim 0.5$ with the rest of the contribution coming from the \texttt{z18}-$Y_e$ model with standard explosion energy with minimal fallback. 
The best-fit CCSN+near-\ch model has essentially no contribution from the SN 1a and is thus identical to the best-fit single CCSN model that corresponds to the \texttt{z19.8}-$S_4$ model with negligible fallback. The PISN involved in the best-fit CCSN+PISN model is from the lightest progenitor corresponding to a He core of $65\,\Msun$ that only produces light and intermediate elements and contributes only to elements from C--Mg with the remaining coming from the \texttt{z10.7}-$Y_e$ model with minor fallback. 

Overall, the 2CCSNe model provides a very good fit and all other scenarios, except single PISN, can also provide good fits to the observed abundance pattern although there is no contribution from near-\ch SN 1a. Our analysis hints towards a pure CCSN origin for this star although some contribution from sub-\ch SN 1a or a low mass PISN cannot be ruled out. For this reason, a clear signature of CCSN cannot be claimed. 

\subsubsection{SMSSJ034249-284215 (Fig.~\ref{fig:groupA_1}b)}
SMSSJ034249-284215 has a metallicity of $\B{Fe}{H}=-1.97$ with 12 elements detected with $Z\leq30$~\citep{reggiani2023AJ} and was analysed in detail in \citet{Jeena_SN_1a_2024}. Although the metallicity is marginally higher than the maximum $\B{Fe}{H}=-2$ for VMP stars, we included this star in our analysis. As noted in \citet{Jeena_SN_1a_2024}, the Al abundance measured by \citet{reggiani2023AJ}  did not account for non-local thermodynamic equilibrium (NLTE) corrections that are typically large for MP stars and give higher values of $\B{Al}{Fe}$ compared to local thermodynamic equilibrium (LTE) analysis \citep{baumueller1997}. For this reason, we consider the observed $\B{Al}{Fe}$ as a lower limit. 

This star has a peculiar pattern for Cr--Mn with sub-solar $\B{Cr}{Mn}$ that none of the sources can fit. Consequently, none of the best-fit models can simultaneously fit Cr and Mn.
The best-fit 2CCSNe model provides the overall best-fit with a $\chi^2=1.6$ resulting from the combination of ejecta from \texttt{z13.7}-$S_4$ model without fallback and the \texttt{z15.2}-$S_4$ model with minimal fallback with standard explosion energy of $1.2\times 10^{51}\,\erg$. 
Except for Cr and Mn, both of which have relatively high deviations of $2.57\sigma$, this model can provide a reasonably good fit for the rest of the elements with minor deviations beyond $1\sigma$ for Ca and Ti. The overall quality of the fit can be classified as acceptable. 
The best-fit CCSN+near-\ch model provides a comparable acceptable fit with $\chi^2=1.7$ where the CCSN ejecta is from the same \texttt{z15.2}-$S_4$ model but with zero fallback. Compared to the best-fit 2CCSNe model, this model can match Ca and Mn but fails to match Na, Si, Ti, and Fe within $1\sigma$ uncertainty.  
The contribution from near-\ch SN 1a in the best-fit model is significant for most elements from Cr--Ni. Compared to the best-fit 2CCSNe model, the quality of fit from the best-fit CCSN+sub-\ch model is worse ($\chi^2=2.2$) with a particularly high deviation for Cr of $3.88\sigma$ but can fit Ti and Ca.  The contribution from sub-\ch SN 1a is significant for elements from Ti--Zn. The overall quality of the fit can be classified as poor particularly due to the large deviation for Cr.
The quality of fit from the best-fit single CCSN model is even worse and can also be classified as poor with $\chi^2=4.72$ with relatively large deviations of $2\hbox{--}4\sigma$ for elements such as Na, Ca, Ti, Cr, Mn, Co and Zn.   
The best-fit CCSN+PISN model also provides a poor fit with $\chi^2=2.64$ with minimal contribution from a PISN resulting from a $130\,\Msun$ He core star and fails to fit Mg, Ca, Ti, Cr, Mn, and Co. Lastly, the best-fit single PISN model gives a very poor fit with a $\chi^2=26.4$ and can be essentially ruled out. 

\begin{figure*}
    \centering
    \includegraphics[width=\columnwidth]{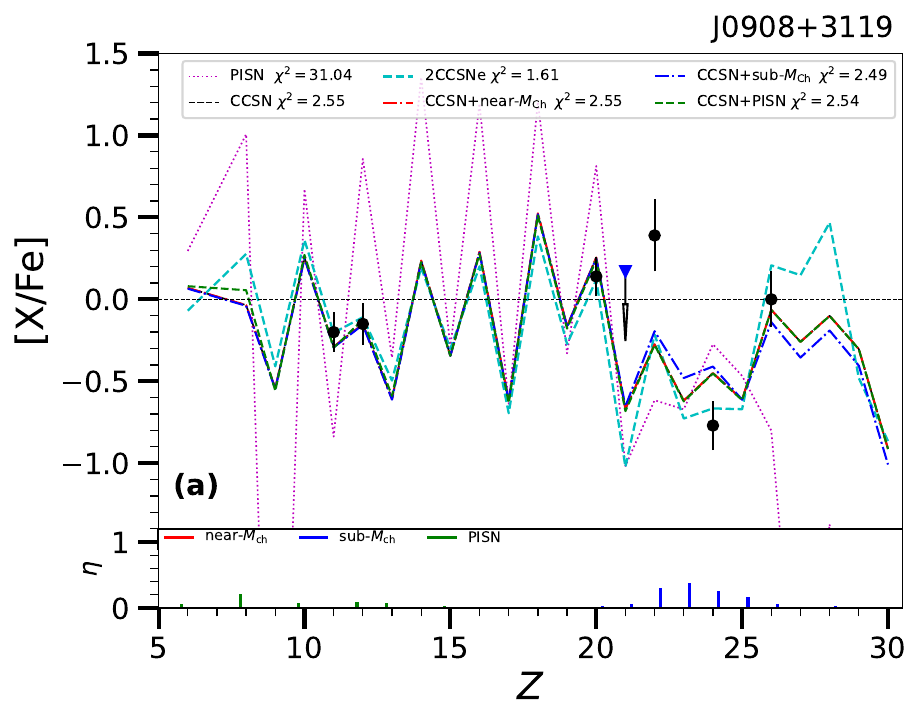}
    \includegraphics[width=\columnwidth]{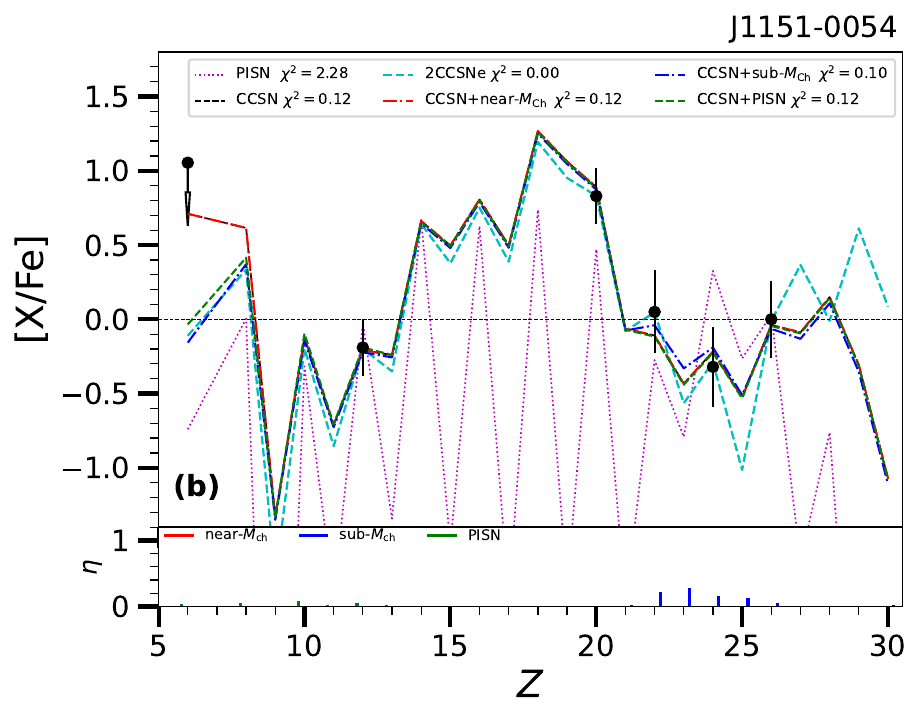}\\
    \includegraphics[width=\columnwidth]{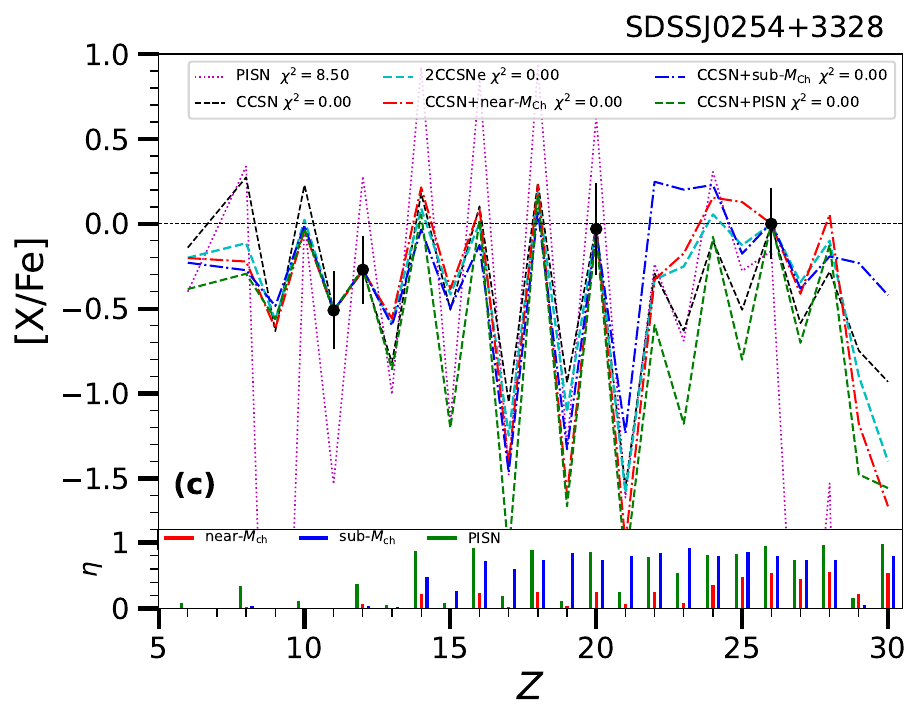}
    \includegraphics[width=\columnwidth]{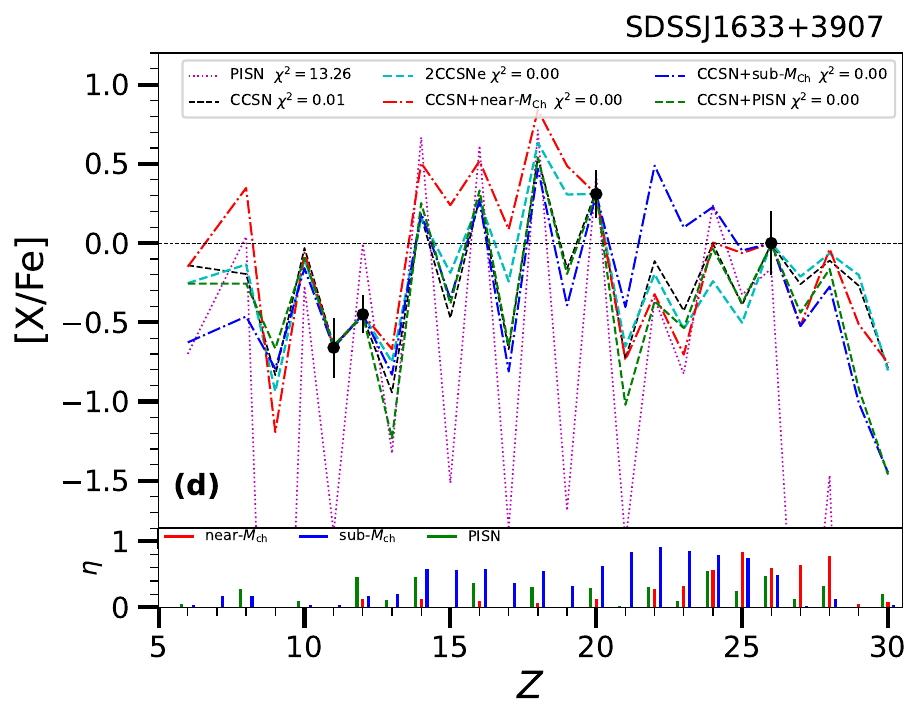}
    \caption{a) Same as Fig.~\ref{fig:groupA_1}, but for J0908+3119~\citep{Li2022ApJ}. b) Same as (a), but for J1151-0054~\citep{Li2022ApJ}. c) Same as (a), but for SDSSJ0254+3328~\citep{Aoki2013AJ}. d) Same as (a), but for SDSSJ1633+3907~\citep{Aoki2013AJ}. Note that the observed Sc is treated as an upper limit in J0908+3119.}
    \label{fig:groupA_2}
\end{figure*}
Overall, due to the peculiar Cr and Mn ratio, none of the scenarios provide a good fit to the observed abundance pattern. 
However, 2CCSNe and CCSN+near-\ch scenarios provide acceptable fits but CCSN+sub-\ch and CCSN+PISN scenarios provide worse fits that can be classified as poor.  The single CCSN scenario also provides a poor fit whereas the single PISN scenario provides a very poor fit. Because none of the scenarios provides a good fit and the quality of fit is comparable from 2CCSNe and CCSN+near-\ch, no clear signature of any source is found.

\subsubsection{HE1207-3108 (Fig.~\ref{fig:groupA_1}c)} \label{sec:ch7_HE1207}
HE1207-3108 has a metallicity of $\B{Fe}{H}=-2.7$ with 11 elements detected with $Z\leq 30$~\citep{yong_norris2013}. 
The best-fit 2CCSNe model provides the overall best fit with $\chi^2=0.60$, which is a combination of ejecta from \texttt{z11.4}-$Y_e$ without fallback and \texttt{z19.6}-$Y_e$ with minimal fallback. This model provides a very good fit where it can match all elements within $1\sigma$ uncertainty except for Mn and Ni with deviations of $1.54\sigma$ and $1.49\sigma$, respectively. The best-fit CCSN+PISN model also provides an equally good fit resulting from the combination of ejecta from  \texttt{z11.4}-$Y_e$ model with low fallback and $75\,\Msun$ He core PISN model. This can also fit all elements within $1\sigma$ uncertainty except for Mn and Ni with slightly lower deviations of $1.57\sigma$ and $1.41\sigma$, respectively.
The best-fit CCSN+near-\ch model also provides a very good fit with $\chi^2=0.67$, where the near-ch SN 1a contributes to all elements from Cr--Ni except for Co. Unlike the best-fit 2CCSNe and CCSN+PISN models, the best-fit model can match Mn and Ni within $1\sigma$ uncertainty, but it fails to match Cr and Fe with deviations of $1.71\sigma$ and $1.27\sigma$, respectively. 
The best-fit CCSN+sub-\ch model provides a slightly worse fit with $\chi^2=0.80$, where it fails to match Si, Cr, Mn, and Ni within the observed $1\sigma$ uncertainty with deviations of $1.06\sigma$, $1.34\sigma$, $1.46\sigma$, and $1.33\sigma$, respectively. The sub-\ch SN 1a only contributes significantly to elements from Sc--V.
The best-fit single CCSN model also provides a slightly worse fit with $\chi^2=0.93$ where it fails to match Si, Mn, and Ni within $1\sigma$ uncertainty with somewhat higher deviations of $1.86\sigma$, $1.26\sigma$, and $1.66\sigma$, respectively.

Overall, 2CCSNe, CCSN+PISN, and CCSN+near-\ch scenarios all provide very good fits to the observed abundance pattern while the best-fit CCSN+sub-\ch provides a good fit. The quality of fit from the single CCSN scenario can be classified as somewhere between acceptable and good. The single PISN scenario provides a very poor fit and can be ruled out. Because multiple scenarios provide a very good fit, no clear signature of any particular source can be claimed. 

\subsubsection{J0025+2305 (Fig.~\ref{fig:groupA_1}d)}
J0025+2305 has a metallicity of $\B{Fe}{H}=-2.81$ with 7 elements detected with $Z\leq30$~\citep{Li2022ApJ}. Although this is an $\alpha$PVMP star with sub-solar \B{Mg}{Fe}, it has super-solar $\B{Ca}{Fe}\sim0.4$. Among CCSN models, such a feature is only found in progenitors that either have O-shell burning or merger of O-shell burning regions with the O-Ne-Mg shell before collapse that results in super solar $\B{X}{Fe}$ for elements from Si--Ca. For this reason, the CCSN models in all the scenarios are exclusively from models that have this feature. 
The best-fit model from 2CCSNe gives the overall best fit with $\chi^2=0.5$ and can match all the elements except for Ti where the deviation is $1.41\sigma$. The best-fit 2CCSNe model is a combination of \texttt{z19.4}-$Y_e$ model without fallback and \texttt{z17.8}-$Y_e$ model with minimal fallback with high explosion energy of $1.2\times10^{52}\,\erg$. Whereas  \texttt{z19.4}-$Y_e$ model undergoes merger of O-burning and O-Ne-Mg shell, \texttt{z17.8}-$Y_e$ model 
 undergoes O-shell burning prior to collapse. 
 
The best-fit CCSN+sub-\ch model provides a very good fit with an almost identical $\chi^2=0.53$ to the best-fit 2CCSNe model. In this case, it is the result of the combination of ejecta from the \texttt{19.6}-$Y_e$ model that undergoes shell merger prior to collapse, and sub-\ch SN 1a from the lightest CO core model of $0.8\,\Msun$ that has elevated intermediate elements. 
In fact, this model can fit all the elements including Ti within the observed $1\sigma$ uncertainty. However, the quality of fit is slightly worse for Cr compared to the best-fit 2CCSNe model. Sub-\ch SN 1a contributes majorly to all elements from Ti--Mn account for $\sim 60\hbox{--}80\%$ of the abundance along with $30\hbox{--}40\%$ contribution for Ca, Sc, and Fe.

The best-fit CCSN+PISN model can also provide a good fit with $\chi^2=0.73$ with Ti being the only outlier with a deviation of $1.79\sigma$. Here, the CCSN model is from \texttt{z21}-$Y_e$ that undergoes shell merger along with the lightest PISN model from a He core of $65\,\Msun$ that only produces elements up to S and primarily contributes to light elements up to Al as evident from the value of $\eta_{\rm PISN}$. The quality of fit from the best-fit single CCSN model is acceptable with $\chi^2=1.43$ where it fails to match  Mg, Ti, Cr, and Fe within $1\sigma$ but can fit all elements within $1.6\sigma$. 
The best-fit CCSN+near-\ch model is effectively the same as the best-fit single CCSN model with near zero contribution from SN 1a. Lastly, the single PISN scenario provides an extremely poor fit with a $\chi^2=42.31$. We note that all CCSN models except for the best-fit CCSN+sub-\ch model have higher explosion energy of $1.2\times 10^{52}\,\erg$ which is due to the elevated Ti found in the star that is naturally produced in such models.

Overall, both 2CCSNe and CCSN+sub-\ch scenarios provide very good fits while the CCSN+PISN scenario provides a good fit. The single CCSN can provide an acceptable fit which is the same as the best-fit CCSN+near-\ch as it has no contribution from near-\ch SN 1a. The single PISN scenario provides a very poor fit and can be ruled out. Because both 2CCSNe and CCSN+sub-\ch scenarios provide very good fits, no clear signature of either CCSN or sub-\ch SN 1a can be claimed. 

\begin{table*}
\centering
{\scriptsize\tabcolsep=3.0pt  % hold it local
%{\tiny
\caption{Same as Table~\ref{tab:best_fit_GA_1}, but for the 4 Group A stars shown in Fig.~\ref{fig:groupA_2}.}
\label{tab:best_fit_GA_2}
\begin{tabular}{|c|c|c|c|c|c|c|c|c|r|}
\hline
    Star&Scenario&Model name&$E_{\rm exp}$ &$\chi^2$&$\alpha$&$\Delta M_{
    \rm cut}$&$\Delta M_{\rm fb}$&$M{\rm_{dil}}$&Outliers\\
    && & ($\times 10^{51}\,\erg$)& &&(\Msun) &(\Msun)& ($\times 10^4\,\Msun$)&\\
    \hline
    \multirow{9}{*}{\rotatebox[origin=c]{90}{J0908+3119}}&PISN&$75\,\Msun$ He core&13.8 &31.04&--&--&--& $2.9\times10^4$&--\\
    \cline{2-10}
    &CCSN&\texttt{z10.9}-$Y_e$&1.2&2.55&--&0.10 &0.07 &8.60&Ti (3.02), Cr (2.13)\\
    \cline{2-10}
    &CCSN+&\texttt{z10.9}-$Y_e+$&1.2&\multirow{2}{*}{2.54} &\multirow{2}{*}{$5\times10^{-4}$} &0.10 &0.07 &8.90& \multirow{2}{*}{Ti (3.04), Cr (2.12)}\\ 
    &PISN&$65 \,\Msun$ He core & 4.9&&& --& --&  $1.8\times10^4$&\\
    \cline{2-10}
    & \multirow{2}{*}{2CCSNe} & \texttt{z10.6}-$Y_e +$&1.2&\multirow{2}{*}{1.61}&\multirow{2}{*}{0.73}&0.00&0.00&$2.5\times10^1$&\multirow{2}{*}{Ti (2.75), Fe (1.22)}\\
    && \texttt{z18}-$Y_e$&1.2& & & 0.38&0.38 &$6.7\times10^1$&\\
    \cline{2-10}
    & CCSN+&\texttt{z10.9}-$Y_e+$&1.2&\multirow{2}{*}{2.55}&\multirow{2}{*}{$10^{-7}$}&0.10 &0.07 &8.63&\multirow{2}{*}{Ti (3.02), Cr (2.13)}\\
     &near-\ch&  N100\_Z0.01& --& &&-- &--&$8.6\times10^7$ &\\ 
     \cline{2-10}
     &CCSN+&\texttt{z10.9}-$Y_e +$&1.2&\multirow{2}{*}{2.49}&\multirow{2}{*}{0.002}&0.10 &0.08 &8.72&\multirow{2}{*}{Ti (2.66), Cr (2.40)}\\
    &sub-\ch&   M09\_05& --&& &-- &--&$4.4\times10^3$ &\\       
       \hline 
       \hline
       \multirow{9}{*}{\rotatebox[origin=c]{90}{J1151-0054}}&PISN&$105\,\Msun$ He core&48.9 &2.28&--&--&--& $8.8\times10^6$&--\\ 
       \cline{2-10}
            &CCSN&\texttt{z16.6}-$Y_e$&1.2&0.12&--&1.86 &1.65 &1.02&None\\
            \cline{2-10}
           &CCSN+&\texttt{z16.6}-$Y_e+$&1.2&\multirow{2}{*}{0.12} &\multirow{2}{*}{$10^{-7}$} & 1.86&0.04 &9.05&\multirow{2}{*}{None}\\
           &PISN&$125 \,\Msun$ He core &78.8 &&&-- &-- &  $9.1\times10^3$&\\
           \cline{2-10}
           & \multirow{2}{*}{2CCSNe} & \texttt{z12.8}-$Y_e +$&12&\multirow{2}{*}{0.00}&\multirow{2}{*}{0.3}&0.00&0.00&$2.4\times10^1$&\multirow{2}{*}{None}\\
           && \texttt{z16.6}-$Y_e$&12& & &0.31 &0.31 &$1.0\times10^1$&\\
           \cline{2-10}
           & CCSN+&\texttt{z16.6}-$Y_e +$&1.2&\multirow{2}{*}{0.12}&\multirow{2}{*}{$10^{-5}$}&1.37 &0.00 &9.24&\multirow{2}{*}{None}\\
           &near-\ch&  N100\_Z0.01&-- & &&-- &-- &$9.2\times10^5$&\\
           \cline{2-10}
           &CCSN+&\texttt{z16.6}-$Y_e +$&1.2&\multirow{2}{*}{0.10}&\multirow{2}{*}{$2\times10^{-3}$}&1.76 &1.55 &1.21&\multirow{2}{*}{None}\\
            &sub-\ch&   M08\_05& --&& & --&-- &$6.0\times10^2$&\\ 
            \hline 
            \hline
     \multirow{9}{*}{\rotatebox[origin=c]{90}{SDSSJ0254+3328}} & PISN& $95\,\Msun$ He core&35.3 &8.50&--&--&--&$2.1\times 10^4$&--\\ 
     \cline{2-10}
       &  CCSN& \texttt{z22}-$S_4$ &1.2&0.00&--&1.83 &1.09 &6.11&None\\
       \cline{2-10}
        &CCSN+&\texttt{z10.1}-$S_4 +$ &0.3&\multirow{2}{*}{0.00} &\multirow{2}{*}{$6\times10^{-4}$} &0.11 &0.06 & 1.10&\multirow{2}{*}{None}\\
        &PISN&$125 \,\Msun$ He core &78.8 &&&-- &-- &$1.8 \times 10^3$&\\
        \cline{2-10}
        & \multirow{2}{*}{2CCSNe} & \texttt{z11.1}-$Y_e +$&0.6& \multirow{2}{*}{0.00}&\multirow{2}{*}{0.25}&0.00&0.00&9.07 &\multirow{2}{*}{None}\\
        & & \texttt{z11.9}-$Y_e $&0.6 & &&0.21 &0.13 &3.02&\\ 
        \cline{2-10}
         & CCSN+&\texttt{z10.5}-$Y_e +$&0.6 &\multirow{2}{*}{0.00}&\multirow{2}{*}{0.02}&0.09 &0.08 &1.43&\multirow{2}{*}{None}\\ 
         &near-\ch&  N100\_Z0.01&-- & &&--&--&$7.0\times10^1$&\\
         \cline{2-10}
         &CCSN+&\texttt{z10.4}-$S_4 +$& 0.3&\multirow{2}{*}{0.00}&\multirow{2}{*}{0.04}&0.10 &0.003 &1.38&\multirow{2}{*}{None}\\
         &sub-\ch&   M09\_10& --&& &--&--&$3.3\times10^1$ &\\
         
         \hline
         \hline
         \multirow{9}{*}{\rotatebox[origin=c]{90}{SDSSJ1633+3907}}&PISN& $100\,\Msun$ He core&41.9 & 13.26& --&--&--&$4.9\times10^4$&--\\
         \cline{2-10}
          &CCSN&\texttt{z11.1}-$S_4 $ &1.2&0.01&--&0.11 &0.04 &2.34&None\\
          \cline{2-10}
         &CCSN+&\texttt{z11.1}-$S_4 +$&0.3& \multirow{2}{*}{0.00}&\multirow{2}{*}{$1\times10^{-3}$} & 0.19&0.07 &3.10&\multirow{2}{*}{None}\\
         &PISN&$120 \,\Msun$ He core +&71.0 &&&-- &-- &$3.1 \times 10^3$&\\
         \cline{2-10}
         & \multirow{2}{*}{2CCSNe} & \texttt{z10.6}-$S_4 +$ &1.2&\multirow{2}{*}{0.00}&\multirow{2}{*}{0.21}&0.00&0.00&9.79&\multirow{2}{*}{None}\\
         &&\texttt{z11.2}-$S_4 $&1.2& & &0.09 &0.03 &2.60&\\
         \cline{2-10}
         & CCSN+&\texttt{z21}-$S_4 +$&1.2&\multirow{2}{*}{0.00}&\multirow{2}{*}{0.12}& 1.34&0.48 &9.76&\multirow{2}{*}{None}\\
          &near-\ch&  N100\_Z0.01& --& && --&--&$7.2\times10^1$&\\
          \cline{2-10}
          &CCSN+&\texttt{z10.1}-$Y_e +$&0.6&\multirow{2}{*}{0.00}&\multirow{2}{*}{0.18}&0.09 &0.06 &3.10&\multirow{2}{*}{None}\\
          &sub-\ch&   M08\_03&-- && & --&--&$1.4\times10^1$ &\\
          \hline
          %&&&&&&&&\\
        \end{tabular}}%end
\end{table*}

\subsubsection{J0908+3119 (Fig.~\ref{fig:groupA_2}a)}
J0908+3119 is an extremely metal-poor (EMP) star with a metallicity of $\B{Fe}{H}=-3.74$ with only 7 elements detected with $Z\leq30$~\citep{Li2022ApJ}. This star has an unusually low $\B{Cr}{Fe}\sim -0.7$ along with very high $\B{Ti}{Cr}\sim1.0$. Such a feature is not found in any of the PISN, CCSN, and SN 1a models. Consequently, none of the best-fit models in any of the scenarios can simultaneously match Ti and Cr.
Nevertheless, the best-fit 2CCSNe model can provide an acceptable fit with $\chi^2=1.61$ resulting from the combination of ejecta from \texttt{z10.6}-$Y_e$ model without fallback with the ejecta from \texttt{z18}-$Y_e$ model with minor fallback. This can fit all elements except Ti and Fe for which the deviations are $2.75\sigma$ and $1.22\sigma$, respectively. Compared to the best-fit 2CCSNe model, the best-fit models from single CCSN, CCSN+near-\ch, and CCSN+sub-\ch all provide a worse fit with large devitations for both Ti and Cr of $\sim2\hbox{--}3\sigma$ and can be classified as poor fits.
The contribution of near-\ch SN 1a and PISN is negligible for the  CCSN+near-\ch and CCSN+PISN scenarios, respectively, and is essentially identical to the single CCSN scenario. There is a minor contribution from sub-\ch SN 1a for Ti--Cr in the best-fit CCSN+sub-\ch model. The single PISN scenario provides a very poor fit with a $\chi^2=31.04$.

Overall, only 2CCSNe provides an acceptable fit while the single CCSN, CCSN+sub-\ch, CCSN+near-\ch, and CCSN+PISN scenarios provide poor fits with negligible or minimal contribution from non-CCSN sources.  The single PISN scenario can be ruled out due to the very poor fit. 
Our analysis indicates that the origin of this star is likely associated with pure CCSN ejecta. This is also the most natural scenario from the point of view of galactic chemical evolution as this star has an extremely low metallicity of $\B{Fe}{H}=-3.74$ which makes SN 1a contribution very unlikely. However, the detection of more elements such as Mn, Co, and Ni could help verify whether pure CCSN ejecta is sufficient or whether SN 1a contribution is required to explain the abundance pattern in this star. Currently, a clear CCSN signature cannot be claimed.

\subsubsection{J1151-0054 (Fig.~\ref{fig:groupA_2}b)}
J1151-0054 is also an EMP star with metallicity $\B{Fe}{H}=-3.51$ and has 5 elements detected with $Z\leq30$~\citep{Li2022ApJ}.
Similar to J0025+2305, this star has highly super-solar $\B{Ca}{Fe}\sim0.8$ and thus can be naturally fit by CCSN models that undergo O-shell burning or merger of O-shell burning regions with the O-Ne-Mg shell. The best-fit 2CCSNe model provides a perfect fit with $\chi^2=0.0$ resulting from a combination of \texttt{z12.8}-$Y_e$ model without fallback and \texttt{z16.6}-$Y_e$ model with a minor fallback where the former undergoes O shell burning and the latter undergoes shell merger. All other models except the single PISN can also provide very good fits. However, all such models effectively correspond to the single CCSN scenario as CCSN contributes nearly $100\,\%$ of all elements in CCSN+near-\ch and CCSN+PISN scenarios, and only $\sim20\,\%$ for Ti--Mn in the CCSN+sub-\ch scenario. 
This implies that similar to J0908+3119, the origin of elements in J1151-0054 is likely associated with pure CCSN ejecta and is consistent with the very low metallicity of this star but more elements are needed to be detected to verify this. Currently, a clear CCSN signature cannot be claimed.

\subsubsection{SDSSJ0254+3328 (Fig.~\ref{fig:groupA_2}c)}
SDSSJ0254+3328 has a metallicity of $\B{Fe}{H}=-2.8$ with only 4 elements detected with $Z\leq30$~\citep{Aoki2013AJ}. Other than the single PISN scenario, the best-fit models from all scenarios can provide perfect fits with $\chi^2=0.0$. Except for the best-fit single CCSN model, all other CCSN models are from low-mass models of $10\hbox{--}12\,\Msun$.  Interestingly, unlike J0908+3119 and J1151-0054, in all scenarios involving CCSN and a non-CCSN source, there is a substantial contribution from the non-CCSN source for multiple elements (see $\eta$ in Fig.~\ref{fig:groupA_2}c). Clearly, more elements are needed to be detected to decipher the most likely source for this star. Because multiple scenarios can provide very good fits, no clear signature of any source can be claimed. 
 
\subsubsection{SDSSJ1633+3907 (Fig.~\ref{fig:groupA_2}d)}
SDSSJ1633+3907 has a metallicity of $\B{Fe}{H}=-2.88$ with only 4 elements detected with $Z\leq30$~\citep{Aoki2013AJ}.
Similar to SDSSJ0254+3328, all scenarios except the single PISN scenario can provide perfect fits with $\chi^2=0.0$. As in the case of J0025+2305, this star also has a super-solar value of \B{Ca}{Fe} of about $0.3$. Consequently, CCSN involved in the best-fit models across all scenarios except for CCSN+sub-\ch undergo O-shell burning before the collapse. The sub-\ch model from the best-fit CCSN+sub-\ch model is from the lowest CO core and He shell mass of $0.8\,\Msun$ and $0.03\,\Msun$, respectively, that naturally produces super-solar values for intermediate elements from Si--Ca. The sub-\ch model accounts for $\gtrsim90\,\%$ of the total abundance for almost all elements from Si--Mn in the best-fit CCSN+sub-\ch model (see Fig.~\ref{fig:groupA_2}d). Overall, the situation is similar to SDSSJ0254+3328 and more elements are needed to be detected for the likely source of SDSSJ1633+3907. Because multiple scenarios can provide very good fits, no clear signature of any source can be claimed.

\subsection{Best-fit Group B Stars}
As per the classification, the best-fit CCSN+near-\ch model has the lowest $\chi^2$ among all the six scenarios. Out of 17 stars, only 2 stars belong to this group. Similar to Group A stars, except for the single PISN scenario, the best-fit models from all other scenarios provide comparable fits to the observed abundance pattern for both stars. The best-fit abundance plot is presented in Fig.~\ref{fig:groupB} with the corresponding details of best-fit models and parameters listed in Table~\ref{tab:best_fit_GB}. The detailed analyses for the 2 stars are presented below.

\subsubsection{HE0553-5340 (Fig.~\ref{fig:groupB}a)}
\begin{figure}
    \centering
     \includegraphics[width=\columnwidth]{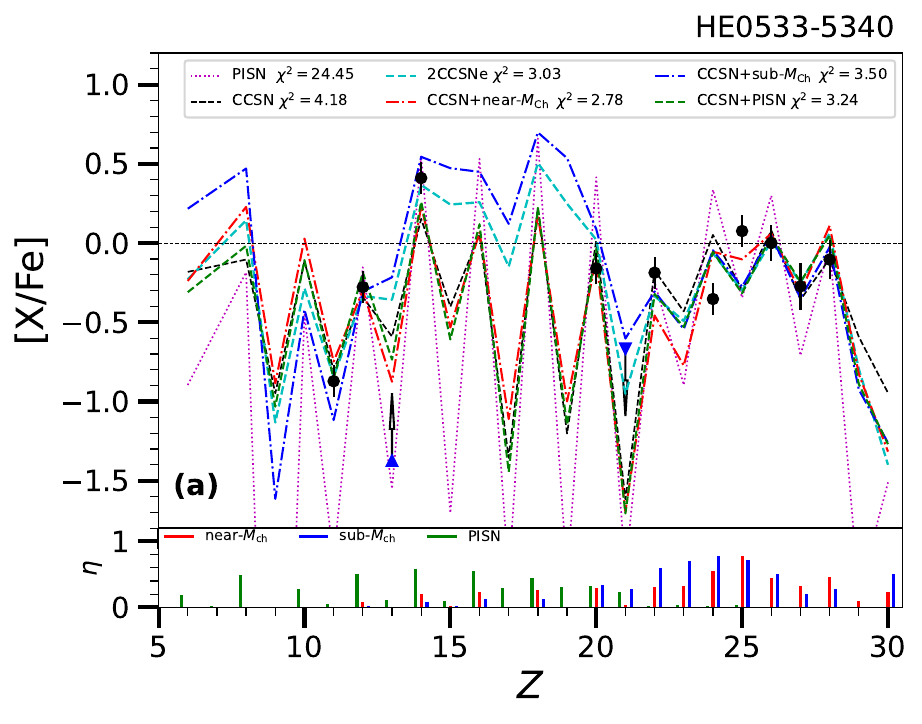}
     \includegraphics[width=\columnwidth]{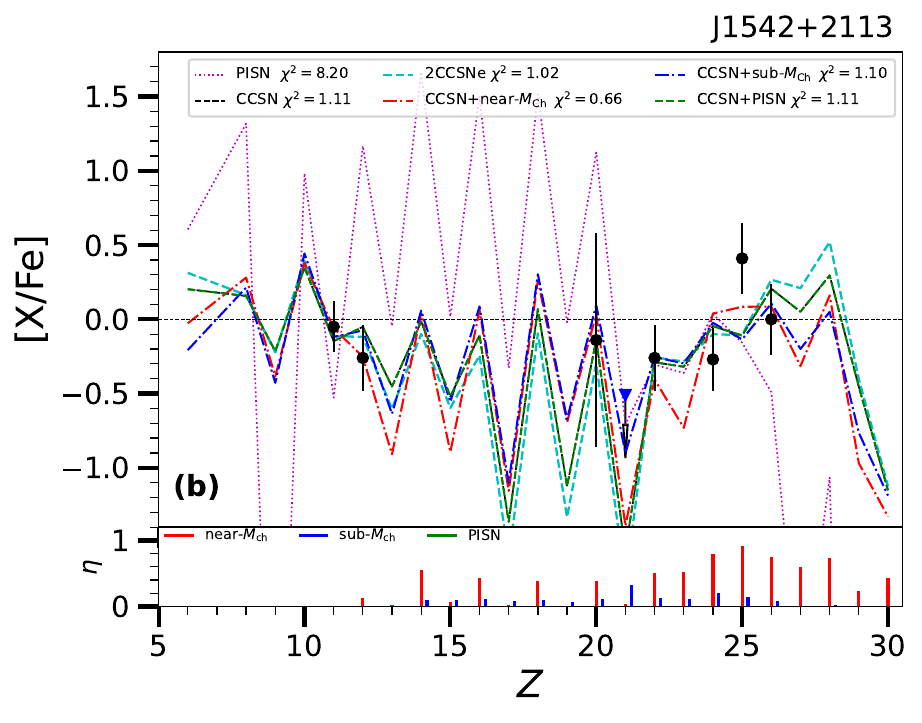}
    \caption{a) Same as Fig.~\ref{fig:groupA_1}, but for star HE0533-5340~\citep{reggiani2023AJ}. b) Same as (a), but for J1542+2113~\citep{Li2022ApJ}. Note that the observed Sc is treated as an upper limit in both stars and the observed Al is treated as a lower limit in HE0533-5340.}
    \label{fig:groupB}
\end{figure}

\begin{table*}
\centering
{\scriptsize\tabcolsep=3.0pt  % hold it local
\caption{Same as Table~\ref{tab:best_fit_GA_1}, but for the 2 Group B stars shown in Fig.~\ref{fig:groupB}. }
\label{tab:best_fit_GB}
\begin{tabular}{|c|c|c|c|c|c|c|c|c|r|}
\hline
    Star&Scenario&Model name&$E_{\rm exp}$ &$\chi^2$&$\alpha$&$\Delta M_{
    \rm cut}$&$\Delta M_{\rm fb}$&$M{\rm_{dil}}$& Outliers\\
    && & ($\times 10^{51}\,\erg$)& &&(\Msun) &(\Msun)& ($\times 10^4\,\Msun$)&\\
          \hline
    \multirow{9}{*}{\rotatebox[origin=c]{90}{HE0533-5340 }}& PISN& $120\,\Msun$ He core& & 24.45& --&-- &--& $3.2\times10^4$& --\\
    \cline{2-10}
          & CCSN&\texttt{z12.2}-$Y_e$&1.2&4.18&-- &0.11 &0.04& 1.19&Si (2.58), Ca (1.77), Cr (4.05), Mn (3.78)\\
          \cline{2-10}
         &CCSN+&\texttt{z11.9}-$Y_e $ +     &1.2   & \multirow{2}{*}{3.24}& \multirow{2}{*}{0.004}&  0.11 &0.06  &1.72&\multirow{2}{*}{Si (1.50), Ca (1.23), Ti (1.34), Cr (2.97), Mn (3.93), Ni (1.21)}\\
         &  PISN                        &$75 \,\Msun$ He core             & & &                                              &--     &--      & $4.3 \times 10^2$& \\
         \cline{2-10}
         &\multirow{2}{*}{2CCSNe}&\texttt{z19.6}-$Y_e +$&1.2& \multirow{2}{*}{3.03}& \multirow{2}{*}{0.13} & 0.00&0.00 & 1.14&\multirow{2}{*}{Ca (1.82), Ti (1.48), Cr (2.89), Mn (3.76), Ni (1.33)}\\ 
         &&\texttt{z10.5}-$Y_e$&1.2 &&& 0.10&0.10 & 7.66&\\ 
         \cline{2-10}
         &  CCSN+ &\texttt{z22}-$Y_e$ &1.2& \multirow{2}{*}{2.78}&  \multirow{2}{*}{0.15}& 0.88&  0.59& 5.56&\multirow{2}{*}{Na (1.25), Si (1.78), Ti (2.70), Cr (2.98), Mn (1.79), Ni (1.71)}\\
         &near-\ch&N100\_Z0.01& --& &&--&--&$3.2\times10^1$ &\\ 
         \cline{2-10}
         &CCSN+ &\texttt{z9.6}-$Y_e$& 1.2& \multirow{2}{*}{3.50}& \multirow{2}{*}{0.05}& 2.11& 1.85& 1.14&\multirow{2}{*}{Na (2.45), Si (1.33), Ca (2.48), Ti (1.18), Cr (3.07), Mn (3.51)}\\
         &sub-\ch& M10\_02& --& & &-- &--&$2.2\times10^1$&\\ 
        \hline
        \hline
    \multirow{9}{*}{\rotatebox[origin=c]{90}{J1542+2115}}&PISN&$75\,\Msun$ He core&13.8 &8.20&--&--&--& $3.1\times10^3$&--\\
    \cline{2-10}
     &CCSN&\texttt{z11.9}-$Y_e$&0.6&1.11&--&0.31 &0.22 &1.80&Cr (1.07), Mn (2.16)\\
     \cline{2-10}
    &CCSN+&\texttt{z11.9}-$Y_e$&0.6&\multirow{2}{*}{1.11} &\multirow{2}{*}{$10^{-7}$} & 0.31&0.22 &1.80&\multirow{2}{*}{Cr (1.07), Mn (2.16)}\\ 
    &PISN&$95\,\Msun$ He core +& 35.3&&& --& --&  $1.8\times10^7$&\\
    \cline{2-10}
    & \multirow{2}{*}{2CCSNe} & \texttt{z10.5}-$Y_e +$&0.6&\multirow{2}{*}{1.02}&\multirow{2}{*}{0.27}&0.00&0.00&4.20&\multirow{2}{*}{Mn (2.15), Fe (1.11)}\\
    && \texttt{z11.2}-$Y_e$&0.6& & &0.20 &0.20 &1.54&\\
    \cline{2-10}
     & CCSN+&\texttt{z18.8}-$Y_e+$&1.2&\multirow{2}{*}{0.66}&\multirow{2}{*}{0.12}&1.27 &1.04 &$1.0\times10^1$&\multirow{2}{*}{Cr (1.47), Mn (1.36)}\\
     &near-\ch&  N100\_Z0.01& --& && --&--&$7.3\times10^1$ &\\
     \cline{2-10}
      &CCSN+&\texttt{z17}-$Y_e +$&1.2&\multirow{2}{*}{1.10}&\multirow{2}{*}{0.03}& 1.94&0.47 &9.80&\multirow{2}{*}{Cr (1.17), Mn (2.27)}\\
    &sub-\ch&   M09\_03&-- && & --&--&$3.2\times10^2$ &\\
            \hline
\end{tabular}
}%end
\end{table*}  

HE0553-5340 has a metallicity of $\B{Fe}{H}=-2.44$ with 12 elements detected with $Z\leq30$~\citep{reggiani2023AJ} and was analysed in detail in~\citet{Jeena_SN_1a_2024}.
Similar to SMSSJ034249-284215, we treat the observed Al as a lower limit in this star as \citet{reggiani2023AJ} did not include NLTE corrections that are typically large for MP stars and yields higher $\B{Al}{Fe}$ compared to LTE analysis \citep{baumueller1997}. 
In this star, the peculiar pattern of the sub-solar value of $\B{Cr}{Fe}$ and super-solar $\B{Mn}{Fe}$ cannot be fit by any of the sources. Consequently, none of the scenarios can match Cr and Mn and fail to provide a good fit. 
The overall best-fit is from the best-fit CCSN+near-\ch model with a $\chi^2=2.78$ which is primarily due to the higher $\B{Mn}{Fe}$ produced by the near-\ch SN 1a model. However, it still fails to match Mn along with Na, Si, Ti, Cr, and Ni with particularly large deviations of $2.70\sigma$ and $2.98\sigma$ for Ti and Cr, respectively. We thus classify the quality  
of fit to be poor. The best-fit 2CCSNe model provides a comparable fit with a $\chi^2=3.03$ but with large deviations of $2.89\sigma$ and $3.76\sigma$ for Cr and Mn, respectively, resulting in an overall poor fit. Compared to the best-fit CCSN+near-\ch model, the best-fit CCSN+sub-\ch and CCSN+PISN models provide slightly worse fits with similarly large deviations for Cr and Mn along with additional outliers. Similar to other stars, the single PISN provides by far the worst fit with $\chi^2=24.45$.   

Overall, although CCSN+near-\ch provides the overall best fit, the quality of fit is poor for all scenarios except for the single PISN scenario for which the quality is very poor. The cause of the poor fit is primarily due to sub-solar $\B{Cr}{Fe}$ and super-solar $\B{Mn}{Fe}$. Because all scenarios provide either a poor or very poor fit, no signature of any source can be claimed. 

\subsubsection{J1542+2115 (Fig.~\ref{fig:groupB}b)}
J1542+2115 has a metallicity of $\B{Fe}{H}=-3.07$ with 8 elements detected with $Z\leq30$~\citep{Li2022ApJ}. The best-fit model from the CCSN+near-\ch scenario provides the overall best fit with $\chi^2=0.66$ from a combination of ejecta from near-\ch and \texttt{z18.8}-$Y_e$ with some fallback. 
The quality of fit can be classified as good as it can match almost all elements within $1\sigma$ except for Cr and Mn where the deviations are $1.47\sigma$ and $1.36\sigma$, respectively, with a substantial contribution from near-\ch SN 1a for most elements from Ti--Zn. 
As in the case of HE0533-5340, the preference for near-\ch SN 1a is due to the super-solar value of $\B{Mn}{Fe}\sim 0.4$.  This is evident from the fact that the best-fit models from single CCSN and 2CCSNe can perfectly match all elements but fail to match Mn with a deviation of up to $2.2\sigma$. In both CCSN and 2CCSNe scenarios, the best-fit models involve low-mass CCSN progenitors ranging from $10.5-11.9\,\Msun$. Compared to the best-fit CCSN+near-\ch model, the quality of fit is slightly worse for these models and can be classified as somewhere between good to acceptable.  The best-fit CCSN+PISN model is effectively the same as the single CCSN model as the contribution of PISN is zero for all elements. The quality of fit for the best-fit CCSN+sub-\ch model is very similar to the single CCSN model with almost identical $\chi^2=1.1$. In this case, however, the CCSN is from an intermediate-mass \texttt{z17}-$Y_e$ model. The contribution of sub-\ch SN 1a is negligible for all elements except for Sc--Mn where the contribution is low but non-negligible with $\eta_{1a}\lesssim 0.2$.

Overall, the CCSN+near-\ch scenario provides a good fit to the observed abundance pattern whereas single CCSN, 2CCSNe, CCSN+sub-\ch, and CCSN+PISN scenarios all provide slightly worse fits which can be classified as somewhere between good and acceptable. The single PISN scenario provides a very poor fit and is ruled out as a possible source. Furthermore, because PISN and sub-\ch SN 1a make negligible contributions to the best-fit CCSN+PISN and CCSN+sub-\ch models, respectively, it indicates that PISN and sub-\ch SN 1a are likely not responsible for the elements observed in this star. Although there are some hints of near-\ch SN 1a signature, 
because 2CCSNe and single CCSN scenarios (along with CCSN+sub-\ch, and CCSN+PISN scenarios) can provide fits which are only slightly worse, no clear signature of near-\ch SN 1a can be claimed. 

\subsection{Best-fit Group C Stars}
\begin{figure*}
    \centering
    \includegraphics[width=\columnwidth]{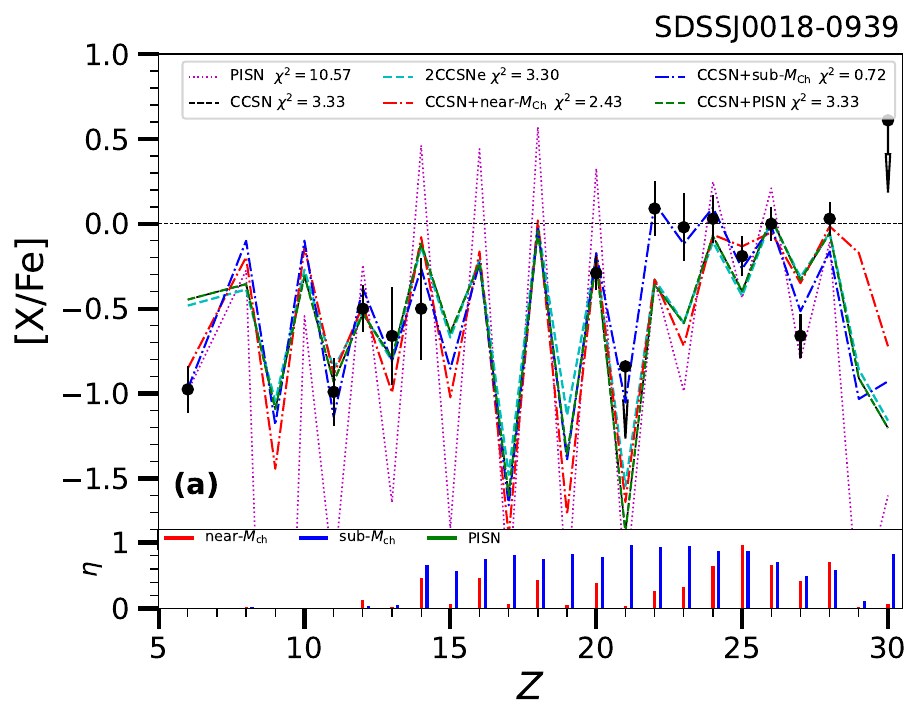}
    \includegraphics[width=\columnwidth]{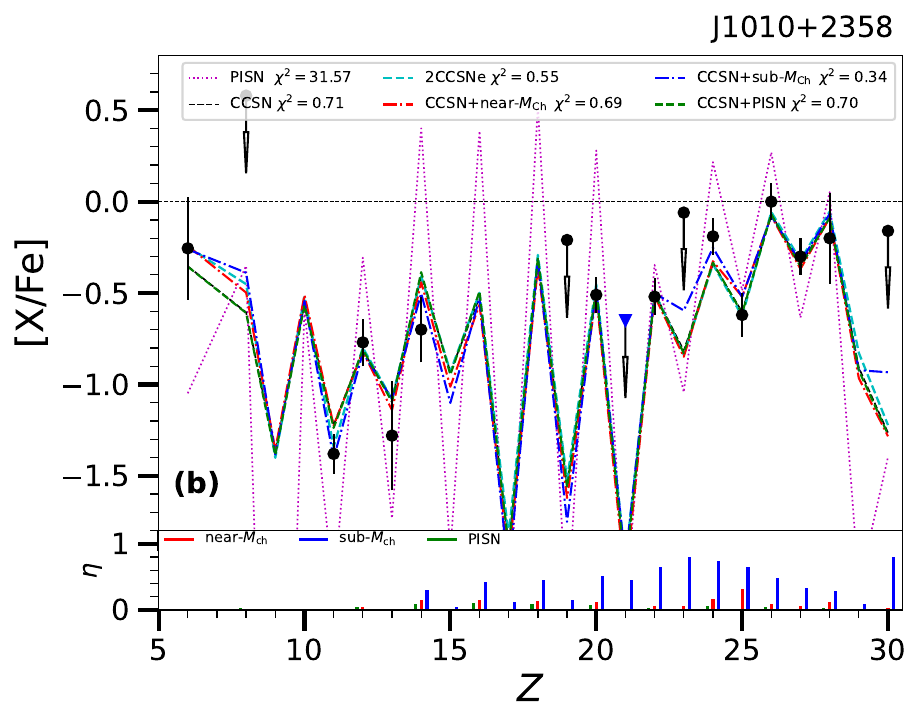}\\
    \includegraphics[width=\columnwidth]{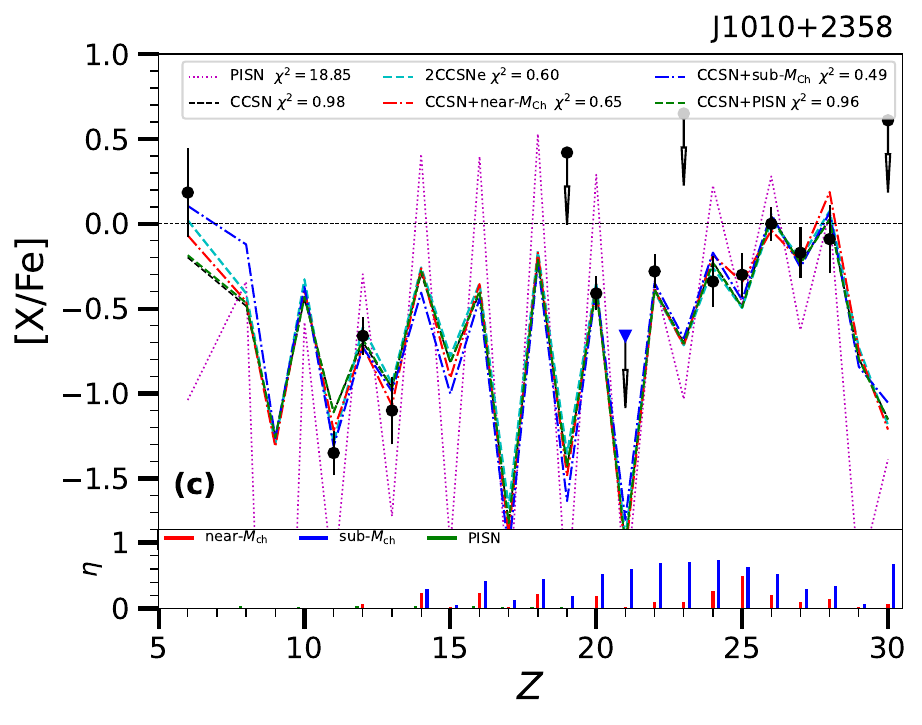}
    \includegraphics[width=\columnwidth]{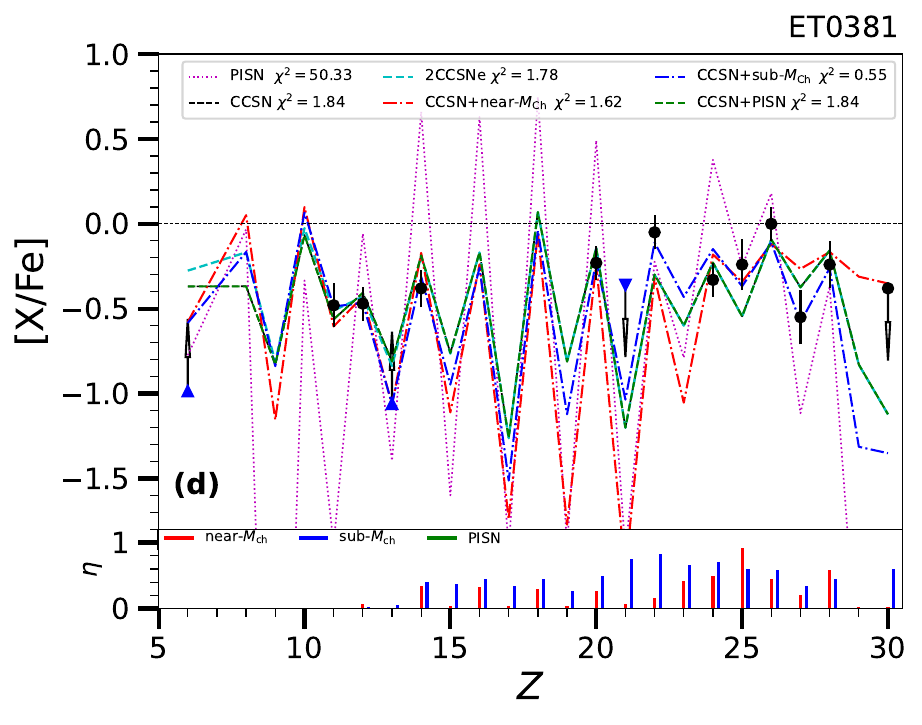}
    \caption{a) Same as Fig.~\ref{fig:groupA_1}, but for star SDSSJ0018-0939~\citep{aoki2014Sci}. b--c) Same as (a), but for star J1010+2358, b) the observed data is from \citetalias{skuladottir2024ApJ}, c) the observed data is from \citetalias{Thibodeaux2024}. d) Same as (a), but for star ET0381~\citep{jabalonka2015}. Note that the observed Sc is treated as an upper limit in J1010+2358 and ET0381 and the observed C and Al are treated as lower limits in ET0381.}
    \label{fig:groupC_1}
\end{figure*}

As per the classification, Group C corresponds to stars where the $\chi^2$ from the best-fit CCSN+sub-\ch model is the lowest. 7 stars belong to this group. The best-fit abundance plots are shown in Fig.~\ref{fig:groupC_1} and Fig.~\ref{fig:groupC_2} with the corresponding information on best-fit models and parameters listed in Table~\ref{tab:best_fit_GC_1} and Table~\ref{tab:best_fit_GC_2}, respectively. Below we discuss each of them in detail. 
\subsubsection{SDSSJ0018-0939 (Fig.~\ref{fig:groupC_1}a)} 

\begin{table*}
\centering
{\scriptsize\tabcolsep=3.0pt  % hold it local
%{\tiny
\caption{Same as Table~\ref{tab:best_fit_GA_1}, but for 4 Group C stars shown in Fig.~\ref{fig:groupC_1}. For J1010+2358, $^*$ and $^\dag$ indicate the observed data is from \citetalias{skuladottir2024ApJ} and \citetalias{Thibodeaux2024}, respectively. }
\label{tab:best_fit_GC_1}
\begin{tabular}{|c|c|c|c|c|c|c|c|c|r|}
\hline
    Star&Scenario&Model name&$E_{\rm exp}$ &$\chi^2$&$\alpha$&$\Delta M_{
    \rm cut}$&$\Delta M_{\rm fb}$&$M{\rm_{dil}}$&Outliers\\
    && & ($\times 10^{51}\,\erg$)& &&(\Msun) &(\Msun)& ($\times 10^4\,\Msun$)&\\
    \hline
    \multirow{9}{*}{\rotatebox[origin=c]{90}{SDSSJ0018-0939}}&  PISN& $120\,\Msun$ He core&71.0&10.57&--&--&--&$4.0\times 10^4$&--\\ 
    \cline{2-10}
       &  CCSN& \texttt{z11.8}-$Y_e$ &1.2&3.33&--&0.11&0.07&1.91&C (3.78), Si (1.28), Ti (2.80), V (2.83), Mn (1.72), Co (2.53)\\
       \cline{2-10}
        & CCSN+&\texttt{z11.8}-$Y_e $ +      &1.2& \multirow{2}{*}{3.33}& \multirow{2}{*}{$10^{-7}$}&0.11 &0.07  &1.90&\multirow{2}{*}{C (3.78), Si (1.28), Ti (2.80), V (2.83), Mn (1.72), Co (2.53)}\\
         &PISN                          &$120 \,\Msun$ He core              &71.0& &                                               &--   &--    &$1.9 \times 10^7$&\\
         \cline{2-10}
         &\multirow{2}{*}{2CCSNe} & \texttt{z11.3}-$Y_e +$&1.2&  \multirow{2}{*}{3.30}& \multirow{2}{*}{0.25}&0.00&0.00&7.85 &\multirow{2}{*}{C (3.52), Si (1.22), Ti (2.61), V (2.82), Mn (1.95), Co (2.66)}\\
         && \texttt{z11.8}-$Y_e $& 1.2& &&0.11& 0.06&2.62& \\ 
         \cline{2-10}
          &CCSN+ &\texttt{z20}-$S_4 +$&12.0 & \multirow{2}{*}{2.43}& \multirow{2}{*}{0.33}&1.07 & 0.00&$1.4\times10^1$&\multirow{2}{*}{Al (1.15), Si (1.41), Ca (1.18), Ti (2.61), V (3.49), Co (2.39)}\\ 
         &near-\ch&  N100\_Z0.01&-- & && --&--&$2.8\times10^1$ &\\ 
         \cline{2-10}
         &CCSN+ &\texttt{z30}-$Y_e +$& 1.2& \multirow{2}{*}{0.72}& \multirow{2}{*}{0.57}&4.23&1.60&$2.5\times10^1$&\multirow{2}{*}{Ca (1.16), Co (1.12), Ni (1.93)}\\
         &sub-\ch&   M10\_05& --&& &-- &--&$1.9\times10^1$ &\\

        \hline
        \hline
     \multirow{9}{*}{\rotatebox[origin=c]{90}{${\rm J1010+2358^*}$}}&  PISN& $125\,\Msun$ He core&78.8&31.57&--&--&--&$4.5\times 10^4$&--\\
     \cline{2-10}
    &CCSN& \texttt{z11.8}-$Y_e$ &1.2&0.71&--& 0.50&0.36 &1.00&Na (1.43), Si (1.62), Cr (1.50) \\
    \cline{2-10}
     &CCSN+& \texttt{z11.8}-$Y_e$+&1.2&\multirow{2}{*}{0.70}&\multirow{2}{*}{$4\times 10^{-5}$}&0.50&0.37&1.00&\multirow{2}{*}{Na (1.29), Si (1.75), Cr (1.44)}\\
         &PISN&$130\,\Msun$ He core&87.3&&&--&--&$2.5 \times 10^4$&\\
         \cline{2-10}
    &\multirow{2}{*}{2CCSNe} & \texttt{z13.5}-$Y_e +$ &1.2&\multirow{2}{*}{0.55}&\multirow{2}{*}{0.08}&0.00&0.00&$1.4\times10^1$&\multirow{2}{*}{Na (1.62), Si (1.52)}\\
    &&\texttt{z11.6}-$Y_e$&1.2& & & 0.39& 0.30& 1.19&\\
    \cline{2-10}
    &CCSN+&\texttt{z11.6}-$Y_e +$&1.2&\multirow{2}{*}{0.69}&\multirow{2}{*}{0.005}& 0.39&0.28 & 1.01& \multirow{2}{*}{Na (1.41), Si (1.44), Cr (1.36)}\\
    &near-\ch&  N100\_Z0.01& --& && --&-- &$2.0\times10^2$&\\ 
    \cline{2-10}
    &CCSN+&\texttt{z15.2}-$Y_e +$& 1.2&\multirow{2}{*}{0.34}&\multirow{2}{*}{0.05}& 1.17&0.97 &2.23&\multirow{2}{*}{Si (1.10)}\\
    &sub-\ch&   M10\_10& --&& &-- &--&$4.2\times10^1$& \\
         \hline
         \hline
   \multirow{9}{*}{\rotatebox[origin=c]{90}{${\rm J1010+2358^\dag}$}}&  PISN& $125\,\Msun$ He core&78.8&18.85&--&--&--&$6.7\times 10^4$&--\\
   \cline{2-10}
    &CCSN& \texttt{z11.8}-$Y_e$ &1.2&0.98&--& 0.50&0.37 &1.04&C (1.47), Na (1.87), Ti (1.04), Mn (1.45)\\
    \cline{2-10}
    &CCSN+& \texttt{z11.8}-$Y_e$+&1.2&\multirow{2}{*}{0.96}&\multirow{2}{*}{$10^{-6}$}&0.50&0.38&1.00&\multirow{2}{*}{C (1.42), Na (1.84), Ti (1.08), Mn (1.48)}\\
         &PISN&$120\,\Msun$ He core&71.0&&&--&--&$2.5 \times 10^5$&\\
         \cline{2-10}
    &\multirow{2}{*}{2CCSNe} & \texttt{z11.9}-$Y_e +$ &1.2&\multirow{2}{*}{0.60}&\multirow{2}{*}{0.11}&0.00&0.00&8.16&\multirow{2}{*}{Ti (1.20), Mn (1.50)}\\
    &&\texttt{z13.4}-$Y_e$&1.2& & & 0.78&0.72 & 1.01&\\
    \cline{2-10}
    &CCSN+&\texttt{z12.1}-$Y_e +$&1.2&\multirow{2}{*}{0.65}&\multirow{2}{*}{0.008}& 0.50&0.41 & 1.02&\multirow{2}{*}{NA (1.02), Ti (1.09), Cr (1.04), Ni (1.39)}\\
    &near-\ch&  N100\_Z0.01&-- & &&-- &-- &$1.3\times10^2$&\\ 
    \cline{2-10}
    &CCSN+&\texttt{z15.2}-$Y_e +$& 1.2&\multirow{2}{*}{0.49}&\multirow{2}{*}{0.03}& 1.17&1.06 &1.45&\multirow{2}{*}{Cr (1.14), MN (1.15)}\\
    &sub-\ch&   M10\_10& --&& &-- &--&$4.7\times10^1$ &\\
     %&&&&&&&&\\
     %&&&&&&&&\\
          \hline
         \hline
     \multirow{9}{*}{\rotatebox[origin=c]{90}{ET0381}}&  PISN& $110\,\Msun$ He core&56.4 &50.33&--&--&-- &$2.19 \times 10^4$&--\\
     \cline{2-10}
     &CCSN& \texttt{z11.3}-$Y_e$ &1.2&1.84&--&  0.29&0.12  &1.18&Si (1.76), Ti (2.49), Cr (1.02), Mn (2.03), Co (1.10) \\
     \cline{2-10}
    &CCSN+& \texttt{z11.3}-$Y_e$+&1.2&\multirow{2}{*}{1.84}&\multirow{2}{*}{$10^{-7}$}&0.29&0.12&1.18&\multirow{2}{*}{Si (1.76), Ti (2.49), Cr (1.02), Mn (2.03), Co (1.10)}\\
         &PISN&$75\,\Msun$ He core&13.8 &&&--&--&$1.18 \times 10^7$&\\
    \cline{2-10}
    &\multirow{2}{*}{2CCSNe} & \texttt{z11.3}-$Y_e +$ &1.2&\multirow{2}{*}{1.78}&\multirow{2}{*}{0.74}&0.00&0.00&2.03&\multirow{2}{*}{Si (1.74), Ti (2.50), Cr (1.01), Mn (2.04), Co (1.09)}\\
    &&\texttt{z19}-$Y_e$&1.2& & & 1.75&1.75 & 5.79&\\
    \cline{2-10}
    &CCSN+&\texttt{z26.5}-$Y_e +$&12&\multirow{2}{*}{1.62}&\multirow{2}{*}{0.18}& 3.09&2.01 & $1.0\times10^1$&\multirow{2}{*}{Si (1.90), Ti (2.67), Cr (1.49), Fe (1.19), Co (1.09)}\\
    &near-\ch&  N100\_Z0.01&-- & &&-- &-- &$4.6\times10^1$&\\ 
    \cline{2-10}
    &CCSN+&\texttt{z17}-$Y_e +$& 1.2&\multirow{2}{*}{0.55}&\multirow{2}{*}{0.17}& 0.97&0.44 &6.00&\multirow{2}{*}{Cr (1.81), Fe (1.10)}\\
    &sub-\ch&   M10\_03& --&& &-- &--&$3.0\times10^1$ &\\         
    \hline
\end{tabular}
}%end
\end{table*}

SDSSJ0018-0939 has a metallicity of $\B{Fe}{H}=-2.5$ with 13 elements detected~\citep{aoki2014Sci} with $Z\leq 30$. A recent study by~\citet{Jeena_SN_1a_2024} showed that this star has a near-smoking gun signature of sub-\ch SN 1a. The key feature in the Ti--Cr region, characterised by \B{X}{Fe}>0 for Ti, V, and Cr, along with \B{Ti}{Cr}>0 along with highly sub-solar $\B{C}{Fe}\sim -1$ was found to be perfectly fit only by the sub-\ch SN 1a model with the CO core mass of $1\,\Msun$. 
We note here that the observed $\log g=5$ rules out any \textit{in situ } depletion of C due to mixing in the observed low-mass star.  
The best-fit CCSN+sub-\ch model is the overall best fit with a $\chi^2=0.72$ and provides a very good fit to the overall abundance pattern with Ni as the only clear outlier with a deviation of $1.93\sigma$ along with Ca and Co being minor outliers with deviations of $1.16\sigma$ and $1.12\sigma$, respectively. The contribution of sub-\ch SN 1a is dominant for most elements from Si--Zn.  
Compared to the CCSN+sub-\ch scenario, the rest of the scenarios provide substantially worse fits. 
The best-fit 2CCSNe provides a poor fit with a $\chi^2=3.30$ as it fails to match C, Si, Ti, V, Mn, and Co with particularly large deviations for C, Ti, V, and Co of $3.52\sigma$, $2.61\sigma$, $2.82\sigma$, and $2.66\sigma$,  respectively. The single CCSN provides a similar poor fit with a $\chi^2=3.33$. The best-fit CCSN+PISN model is effectively the same as the best-fit single CCSN models as there is no contribution from PISN. The best-fit CCSN+near-\ch model provides a slightly better fit compared to the 2CCSNe scenario with a $\chi^2=2.43$ as it can fit C. However, the quality of the overall fit is poor as it fails to match Al, Si, Ca, Ti, V, and Co with high deviations for Ti, V, and Co of $2.61\sigma$, $3.49\sigma$, and $2.39\sigma$, respectively. The single PISN scenario provides a very poor fit with a $\chi^2=10.57$.         

Overall, while CCSN+sub-\ch scenario provides a very good fit to the observed abundance pattern, all other scenarios provide poor fits except for the single PISN for which the fit is very poor. This star thus has a clear signature of sub-\ch SN 1a that is primarily due to the unique abundance feature from Ti--Cr along with highly sub-solar $\B{C}{Fe}$. 

\subsubsection{J1010+2358 (Fig.~\ref{fig:groupC_1}b-c)}
J1010+2358 has a metallicity of $\B{Fe}{H}\sim -2.5$ with 13 elements detected with $Z\leq30$~\citepalias{skuladottir2024ApJ,Thibodeaux2024}. This star was originally identified as the first ever VMP star with a clear signature of PISN~\citep{xing2023Natur}. However, \citet{jeena_CCSN2024} have found that in addition to single PISN, the observed abundance pattern can also be fit perfectly by low mass CCSN models of both Pop III and Pop II stars.
They pointed out that key elements such as C, O, and Al needed to be detected in order to distinguish between PISN and CCSN. 
Following this, \citetalias{Thibodeaux2024} and \citetalias{skuladottir2024ApJ} independently measured the elemental abundance in J1010+2358 using new high-resolution spectra from Keck/HIRES and VLT/UVES, respectively, where they were able to measure the critical missing elements C and Al along with the detection of Na and Sc that clearly ruled out PISN as the possible source. 
Based on the new measurements, we reanalysed the abundance of J1010+2358 for all the six scenarios which are presented in detail in \citet{jeena_LAMOST_revisit_2024} and shown here in Fig.~\ref{fig:groupC_1}b--c. Similar to \citetalias{Thibodeaux2024} and \citetalias{skuladottir2024ApJ}, we also find that PISN provides an extremely poor fit to the newly observed abundance patterns and can be ruled out as a possible source (see Fig.~\ref{fig:groupC_1}b--c). 
We find that other than the single PISN scenario, which provides a very poor fit, all other scenarios can match the overall abundance pattern very well.
The best-fit CCSN+sub-\ch model provides the overall best-fit with $\chi^2 =0.34$ and $\chi^2=0.49$ for the data from \citetalias{skuladottir2024ApJ} and \citetalias{Thibodeaux2024}, respectively, and can match the abundances of almost all the observed elements within $1\sigma$ uncertainty. For the data from \citetalias{skuladottir2024ApJ}, only Si is a minor outlier with a deviation of $1.10\sigma$ whereas for the data from \citetalias{Thibodeaux2024}, Cr and Mn are minor outliers with deviations of $1.14\sigma$ and $1.15\sigma$, respectively.
In both cases, the best-fit result is from the combination of ejecta from \texttt{z15.2}-$Y_e$ with some fallback, and the sub-\ch SN 1a model M10\_10. Importantly, the best-fit 2CCSNe models can also provide an equally good fit and can match almost all elements with Na and Si being minor outliers with deviations of $1.62\sigma$ and $1.52\sigma$, respectively, for the data from \citetalias{skuladottir2024ApJ} and Ti and Mn being minor outliers with deviations of $1.20\sigma$ and $1.50\sigma$, respectively, for the data from \citetalias{Thibodeaux2024}. 
The best-fit single CCSN models provide a very good fit with a slightly higher $\chi^2$ and can match most elements with three outliers for both detections whose deviations range from $\sim 1.01\hbox{--}1.90\sigma$. The best-fit CCSN+near-\ch models can also provide similar good fits with only 3-4 minor outliers but the contribution from SN 1a is negligible for all elements except Mn. Lastly, best-fit CCSN+PISN models provide a fit identical to single CCSN models with nearly zero contribution from PISN for any of the elements.     

Overall, although the observed abundance pattern from both detections in J1010+2358 can be best fit by the CCSN+sub-\ch scenario, all scenarios except the single PISN, provide very good fits and no clear signature of any source can be claimed. Because the best-fit CCSN+PISN models have essentially zero contribution from PISN and the single PISN scenario provides a very poor fit, it indicates that there is not even the slightest hint of a PISN feature in the two newly observed abundance patterns. 

\subsubsection{ET0381 (Fig.~\ref{fig:groupC_1}d)}\label{sec:ch7_ET0381}
ET0381 has a metallicity of $\B{Fe}{H}=-2.44$ with 13 elements detected with $Z\leq30$ and belongs to the Sculptor dwarf galaxy~\citep{jabalonka2015}. Because this star has a low $\log g=1.15$, the initial C in the star would have been considerably depleted corresponding to a correction of $\Delta \B{C}{Fe}\sim+0.8$~\citep{Placco2014Apj}. For this reason, we
treat the observed value of C in this star as a lower limit. We also treat the observed value of Al as a lower limit similar to SMSSJ034249-284215 and HE0533-5340 as it does not account for NLTE corrections. 
The best-fit CCSN+sub-\ch model provides the overall best fit with $\chi^2=0.55$ and provides a very good fit that can match all elements except Cr which is the only clear outlier with a deviation of $1.81\sigma$ and Fe being a very minor outlier with a deviation of $1.1\sigma$. The best-fit CCSN+sub-\ch model is a combination of \texttt{z17}-$Y_e$ CCSN model with minimal fallback and sub-\ch SN 1a model M10\_03 where the latter contributes substantially to all the elements from Si--Zn. 
The best-fit CCSN+near-\ch model also provides a decent fit with $\chi^2=1.62$ where near-\ch SN 1a contributes to most of the elements above Si with considerable contribution from  Cr--Ni. In this case, it fails to match the abundance of Si, Ti, Cr, Fe, and Co within the $1\sigma$ uncertainty with deviations of $1.90\sigma$, $2.67\sigma$, $1.49\sigma$, $1.19\sigma$, and $1.09\sigma$, respectively. The quality of the fit can be considered to be good except for Ti. The situation is quite similar for both the single CCSN and 2CCSNe scenarios, with a marginally higher $\chi^2$ of $1.84$ and $1.78$, respectively. The quality of fit is also comparable to the best-fit  CCSN+near-\ch model where there are five outliers with similar deviations with the highest deviation for Ti. The best-fit CCSN+PISN model is effectively the same as the best-fit single CCSN model as there is zero contribution from PISN. The best-fit single PISN model provides an extremely poor fit with $\chi^2=50.33$ and can be ruled out. 

Overall, the best-fit CCSN+sub-\ch model provides a very good fit to the observed abundance pattern whereas all other scenarios except single PISN provide fits which can be considered to be somewhere between acceptable and good. Thus, this star potentially can be considered to have a somewhat clear signature of sub-\ch SN 1a. 
It is important to note, however, that the reason for the best-fit CCSN+sub-\ch model being clearly better than other best-fit models is due to its ability to fit Ti. However, the TiII abundance adopted for this star that is recommended by \citet{jabalonka2015} is $0.41$ dex higher than the TiI abundance. A lower Ti abundance will clearly improve the quality of fit from other scenarios and impact the association of this star with sub-\ch SN 1a. Additionally, unlike SDSSJ0018-0939, the highly sub-solar $\B{C}{Fe}$ in ET0381 is due to internal depletion and consequently does not provide independent confirmation of SN 1a contribution.  We also note that compared to SDSSJ0018-0939, where all other scenarios provide a poor fit, the fit from other scenarios is reasonable.  Overall, we conclude that although the abundance pattern observed in ET0381 has a reasonably clear sub-\ch SN 1a signature, the signature is not as strong as SDSSJ0018-0939. In this regard, the detection of V along with a robust prediction for Ti abundance could help to clarify the situation. 

\begin{figure*}
    \centering
    \includegraphics[width=\columnwidth]{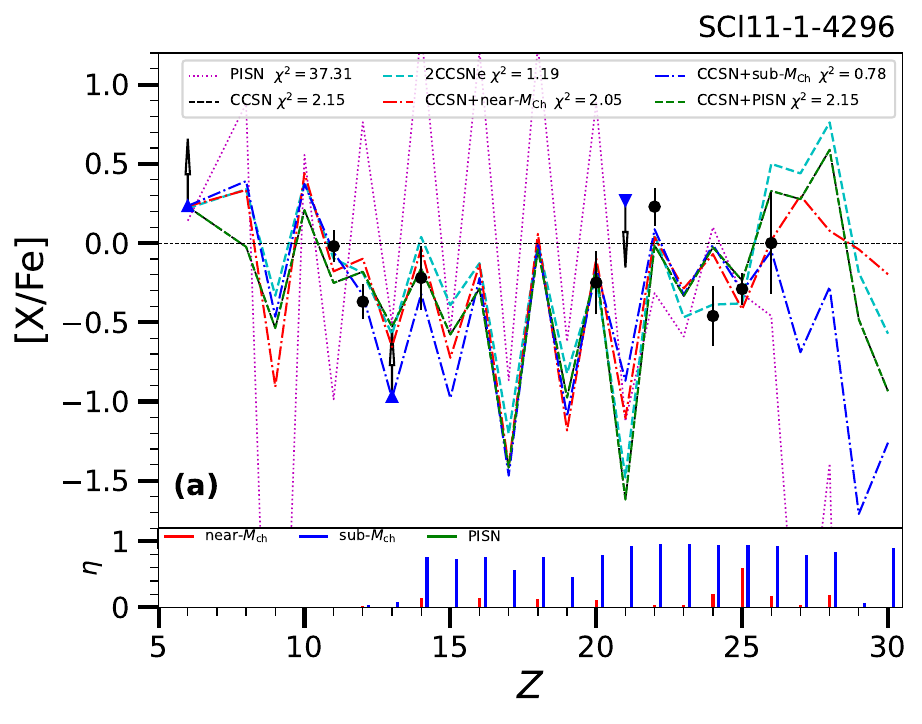}
    \includegraphics[width=\columnwidth]{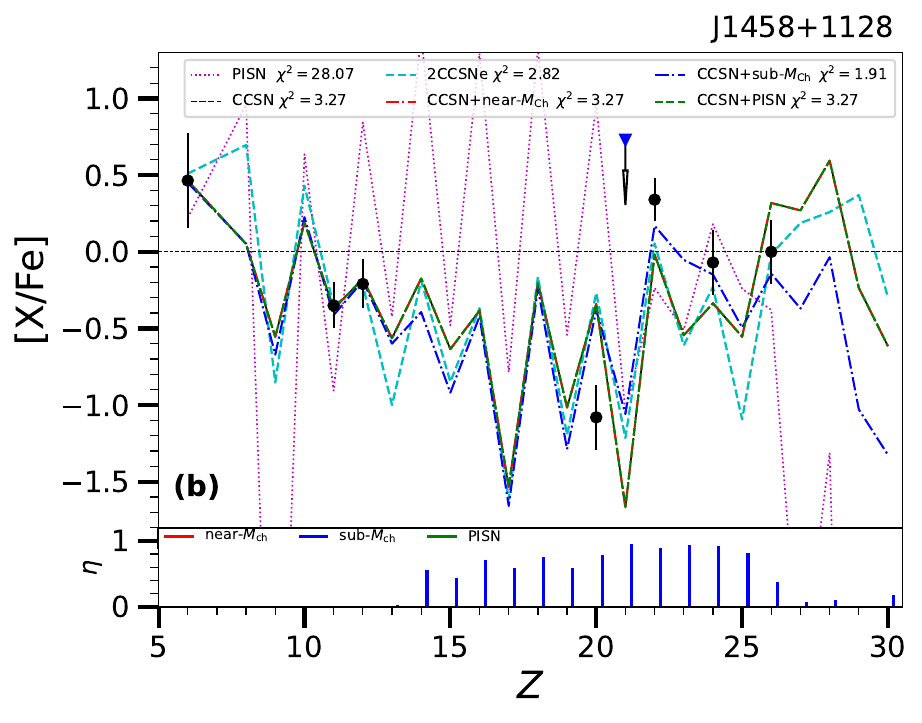}\\
    \includegraphics[width=\columnwidth]{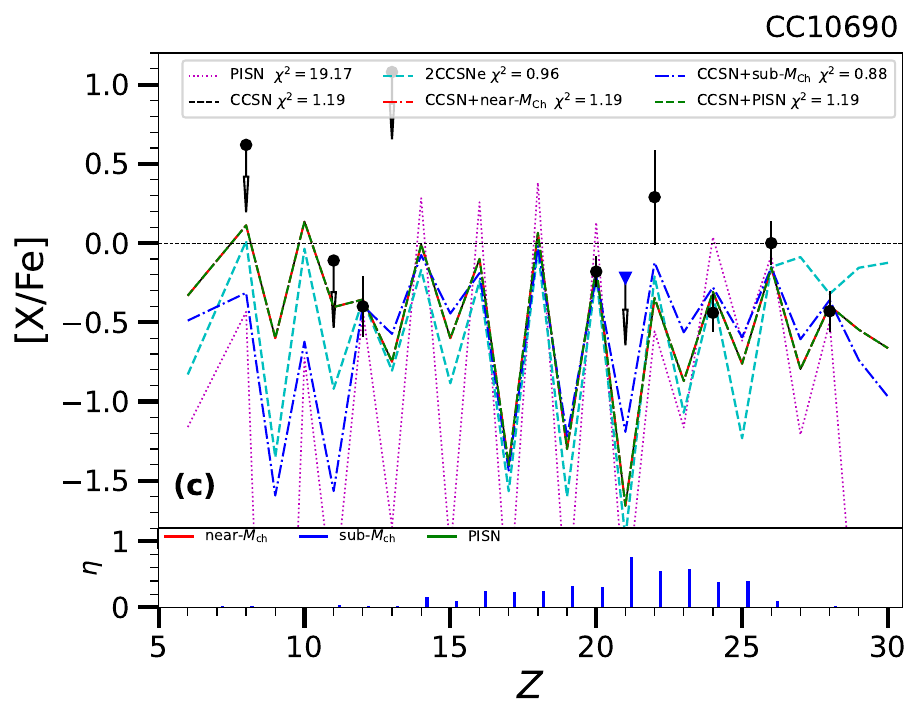}
    \includegraphics[width=\columnwidth]{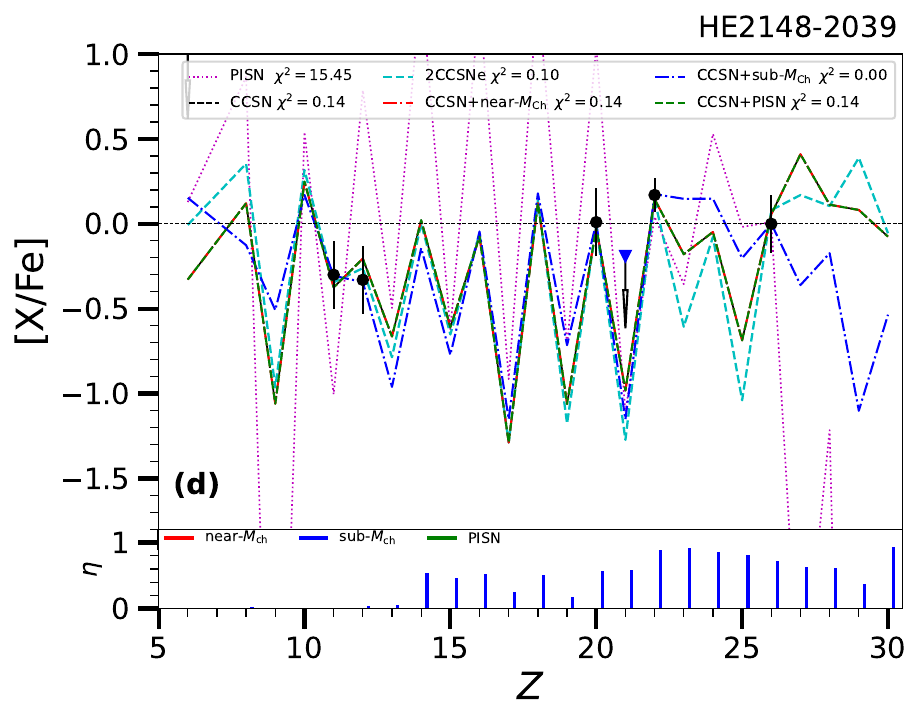}
    \caption{a) Same as Fig.~\ref{fig:groupA_1}, but for star Scl11\_1\_4296~\citep{simon2015ApJ}. b) same as a) but for star J1458+1128~\citep{Li2022ApJ}. c) same as a) but for star CC10690~\citep{norris2017ApJS}. d) same as a) but for star HE2148-2039~\citep{purandardas2021ApJ}.
    Note that the observed Sc is treated as an upper limit in all stars and C and Al are treated as lower limits in Scl11\_1\_4296 (see Sec.~\ref{sec:ch7_Scl11}).}
    \label{fig:groupC_2}
\end{figure*}

\begin{table*}
\centering
{\scriptsize\tabcolsep=3.0pt  % hold it local
%{\tiny
\caption{Same as Table~\ref{tab:best_fit_GA_1}, but for 4 Group C stars shown in Fig.~\ref{fig:groupC_2}.}
\label{tab:best_fit_GC_2}
\begin{tabular}{|c|c|c|c|c|c|c|c|c|r|}
\hline
    Star&Scenario&Model name&$E_{\rm exp}$ &$\chi^2$&$\alpha$&$\Delta M_{
    \rm cut}$&$\Delta M_{\rm fb}$&$M{\rm_{dil}}$&Outliers\\
    && & ($\times 10^{51}\,\erg$)& &&(\Msun) &(\Msun)& ($\times 10^4\,\Msun$)&\\
    \hline
     \multirow{9}{*}{\rotatebox[origin=c]{90}{Scl11\_1\_4296}}&PISN&$85\,\Msun$ He core& 23.2&37.31&--&--&--& $5.0\times10^4$&--\\
     \cline{2-10}
     &CCSN&\texttt{z10.5}-$Y_e$&1.2&2.15&--& 0.19&0.14 &4.57&Na (2.31), Mg (1.72), Ti (2.04), Cr (2.28), Fe (1.03) \\
     \cline{2-10}
    &CCSN+&\texttt{z10.5}-$Y_e$+&1.2&\multirow{2}{*}{2.15} & \multirow{2}{*}{$10^{-7}$}& 0.19&0.14 &4.57&\multirow{2}{*}{Na (2.31), Mg (1.72), Ti (2.04), Cr (2.28), Fe (1.03)}\\ 
    &PISN&$90\,\Msun$ He core&28.9 &&& --& --&  $4.57\times10^7$&\\
    \cline{2-10}
    & \multirow{2}{*}{2CCSNe} & \texttt{z10.6}-$Y_e +$&1.2&1.19&0.77&0.00&0.00&$1.3\times10^1$&\multirow{2}{*}{Mg (1.65), Si (1.29), Ti (1.48), Fe (1.56)}\\
    && \texttt{z18.8}-$Y_e$&1.2& & & 1.46&1.46 &$4.4\times10^1$&\\
    \cline{2-10}
     & CCSN+&\texttt{z17}-$Y_e+$&12&2.05&0.01& 1.55&1.29 &$1.8\times10^1$&\multirow{2}{*}{Na (1.58), Mg (2.47), Ti (1.60), Cr (2.05), Mn (1.27)}\\
     &near-\ch&  N100\_Z0.01&-- & && --&--&$1.78\times 10^3$&\\
     \cline{2-10}
      &CCSN+&\texttt{z18.8}-$Y_e +$&1.2&0.78&0.07& 1.65&1.62 &$2.6\times10^1$&\multirow{2}{*}{Ti (1.19), Cr (2.36)}\\
    &sub-\ch&   M10\_03&-- && & --&--&$3.4\times10^2$ &\\
    \hline
    \hline
      \multirow{9}{*}{\rotatebox[origin=c]{90}{J1458+1128}}&PISN&$85\,\Msun$ He core&23.2 &28.07&--&--&--& $2.7\times10^4$& --\\
      \cline{2-10}
     &CCSN&\texttt{z10.4}-$Y_e$&1.2&3.27&--&0.21 &0.18 &1.60&Ca (3.52), Ti (2.56), Cr (1.27), Fe (1.52)\\
     \cline{2-10}
    &CCSN+&\texttt{z10.4}-$Y_e$&1.2& \multirow{2}{*}{3.27}&\multirow{2}{*}{$10^{-7}$} & 0.21&0.18 &1.60& \multirow{2}{*}{Ca (3.52), Ti (2.56), Cr (1.27), Fe (1.52)}\\ 
    &PISN&$75\,\Msun$ He core +& 13.8&&& --& --&  $1.6\times10^7$&\\
    \cline{2-10}
    & \multirow{2}{*}{2CCSNe} & \texttt{z12.5}-$Y_e +$&12&\multirow{2}{*}{2.82}&\multirow{2}{*}{0.17}&0.00&0.00&$6.3\times10^1$&\multirow{2}{*}{Ca (3.87), Ti (2.04)}\\
    && \texttt{z26}-$Y_e$&12& & &3.94 &3.94 &$1.3\times10^1$&\\
    \cline{2-10}
    & CCSN+&\texttt{z10.4}-$Y_e +$&1.2&\multirow{2}{*}{3.27}&\multirow{2}{*}{$10^{-7}$}&0.21 &0.18 &1.60&\multirow{2}{*}{Ca (3.52), Ti (2.56), Cr (1.27), Fe (1.52)}\\
     &near-\ch&  N100\_Z0.01&-- & && --&--&$1.6\times10^7$ &\\
     \cline{2-10}
    &CCSN+&\texttt{z10.6}-$Y_e +$&1.2&\multirow{2}{*}{1.91}&\multirow{2}{*}{0.004}&0.20 &0.19 &1.90&\multirow{2}{*}{Ca (3.33), Ti (1.22)}\\
    &sub-\ch&   M09\_05&-- && & --&--&$4.6\times10^2$ &\\ 
    \cline{2-10}
    \hline
    \hline
    \multirow{9}{*}{\rotatebox[origin=c]{90}{CC10690}}&PISN&$115\,\Msun$ He core&63.6 &19.17&--&--&--& $1.8\times10^4$&--\\
    \cline{2-10}
     &CCSN&\texttt{z18.4}-$S_4$&1.2&1.19&--&0.79 &0.00 &1.50& Ti (2.12), Cr (1.10), Fe (1.11) \\
     \cline{2-10}
      &CCSN+&\texttt{z18.4}-$S_4$&1.2&\multirow{2}{*}{ 1.19}&\multirow{2}{*}{$10^{-7}$} & 0.79&0.00 &1.50&\multirow{2}{*}{Ti (2.12), Cr (1.10), Fe (1.11)}\\
      &PISN&$105\,\Msun$ He core +&48.9 &&& & &  $1.5\times10^7$&\\
      \cline{2-10}
    & \multirow{2}{*}{2CCSNe} & \texttt{z27}-$Y_e +$&12&\multirow{2}{*}{0.96}&\multirow{2}{*}{0.55}&0.00&0.00&$1.0\times10^1$& \multirow{2}{*}{Ti (1.67), Fe (1.08)}\\
    && \texttt{z29}-$Y_e$&12& & & 6.43&1.09 &$1.2\times10^1$&\\
    \cline{2-10}
     & CCSN+&\texttt{z18.4}-$S_4 +$&1.2&\multirow{2}{*}{1.19}&\multirow{2}{*}{$10^{-7}$}& 0.79&0.00 &1.50&\multirow{2}{*}{Ti (2.12), Cr (1.10), Fe (1.11)}\\
     &near-\ch&  N100\_Z0.01&-- & && --&--&$1.5\times10^7$ &\\
     \cline{2-10}
      &CCSN+&\texttt{z13.7}-$S_4 +$&1.2&\multirow{2}{*}{0.88}&\multirow{2}{*}{0.05}& 0.19&0.02 &1.10&\multirow{2}{*}{Ti (1.38), Cr (1.10), Fe (1.10)}\\
    &sub-\ch&   M08\_05&-- && & --&--&$2.1\times10^1$ &\\

    \hline
    \hline
    \multirow{9}{*}{\rotatebox[origin=c]{90}{HE2148-2039}}&PISN&$90\,\Msun$ He core&28.9 &15.45&--&--&--& $1.8\times10^4$&--\\
    \cline{2-10}
      &CCSN&\texttt{z17}-$Y_e$&12&0.14&--& 1.16& 0.14&$2.4\times10^1$&None\\
      \cline{2-10}
     &CCSN+&\texttt{z17}-$Y_e+$&12& \multirow{2}{*}{0.14}&\multirow{2}{*}{$10^{-7}$} &1.16 &1.09 &$2.4\times10^1$&\multirow{2}{*}{None}\\
     &PISN&$90 \,\Msun$ He core &28.9 &&& --& --&  $2.4\times10^4$&\\
     \cline{2-10}
    & \multirow{2}{*}{2CCSNe} & \texttt{z16.8}-$Y_e +$&12&\multirow{2}{*}{ 0.10}&\multirow{2}{*}{0.46}& 0.00&0.00&$3.8\times10^1$&\multirow{2}{*}{None}\\
    && \texttt{z26}-$Y_e$&12& & &3.32 &3.22 &$3.2\times10^1$&\\
    \cline{2-10}
     & CCSN+&\texttt{z17}-$Y_e+$&12&\multirow{2}{*}{0.14}&\multirow{2}{*}{$10^{-7}$}&1.16 &0.14 &$2.4\times10^1$&\multirow{2}{*}{None}\\
     &near-\ch&  N100\_Z0.01&-- & && --&--&$2.4\times10^9$ &\\ 
     \cline{2-10}
    &CCSN+&\texttt{z10.9}-$Y_e +$&0.6&\multirow{2}{*}{0.00}&\multirow{2}{*}{0.02}& 0.20& 0.18&2.31&\multirow{2}{*}{None}\\
    &sub-\ch&   M09\_10&-- && &-- &--&$1.1\times10^2$ &\\ 
            \hline 
\end{tabular}
}%end
\end{table*}  

\subsubsection{Scl11\_1\_4296 (Fig.~\ref{fig:groupC_2}a)}\label{sec:ch7_Scl11}

Scl11\_1\_4296 is an EMP star with a metallicity $\B{Fe}{H}=-3.77$ with 11 elements detected with $Z\leq30$ and belongs to the Sculptor dwarf galaxy~\citep{simon2015ApJ}. This star has a somewhat peculiar Na--Mg pattern with a solar value of $\B{Na}{Fe}$ but with a super-solar value of $\B{Na}{Mg}$ along with sub-solar $\B{Cr}{Mn}$ which is difficult to match. Because this star has a low $\log g=1.45$, we treat the observed C abundance as a lower limit as it is likely depleted during the evolution of the star~\citep{Placco2014Apj}. Similar to ET0381, we also treat the observed value of Al as a lower limit due to the lack of NTLE corrections.
The best-fit model from the CCSN+sub-\ch scenario provides the overall best fit with $\chi^2=0.78$ with an overall good fit that can match all the elements within $1\sigma$ uncertainty except Cr which has a deviation of $2.36\sigma$ and Ti being a minor outlier with a deviation of $1.19\sigma$. The best-fit model is a combination of ejecta from \texttt{z18.8}-$Y_e$ model with a large fallback and sub-\ch SN 1a model M10\_03, where the former contributes to elements up to Al and the latter contributes to the rest of the elements. The super-solar value of $\B{Ti}{Fe}\sim 0.25$ can only be matched by the sub-\ch SN 1a model and it contributes to $\gtrsim80\,\%$ to all elements from Si--Ni (see $\eta$ in Fig.~\ref{fig:groupC_2}a) with $\sim 93\%$ of Fe coming from SN 1a. However, the highly super-solar $\B{Sc}{Fe}\sim 0.27$, which is used as an upper limit in our analysis, is a problem for the best-fit model. This is because super-solar $\B{Sc}{Fe}$ can only be produced in CCSN models via neutrino-processed material or in shell merger models. However, in the best-fit model,  Fe contribution from CCSN is only $\sim 7\%$ which would require the CCSN model to have an unusually high $\B{Sc}{Fe}\sim 1.4$ in order to fit Sc. Such high $\B{Sc}{Fe}$ is very unlikely even when accounting for neutrino-induced nucleosynthesis. This indicates that although the CCSN+sub-\ch scenario does provide a good fit when neglecting Sc, a larger contribution from CCSN would be required to be consistent with the observed Sc. On the other hand, if extreme shell merger models that have high $\B{Sc}{Fe}$ are invoked, then they would also have highly super-solar $\B{X}{Fe}$ for Si and Ca that will severely affect the quality of fit.

The best-fit 2CCSNe model provides a slightly worse fit with $\chi^2=1.19$ where it can match Cr perfectly but fails to match  Mg, Si, Ti and Fe within $1\sigma$ uncertainty with deviations of $1.65\sigma$, $1.29\sigma$, $1.48\sigma$, and $1.56\sigma$, respectively. Although the number of outliers is high compared to the total number of observed elements, because their deviations are mostly $\lesssim 1.6\sigma$,  we classify this as an acceptable fit.  
Compared to the best-fit 2CCSNe model, the best-fit CCSN+near-\ch model provides a considerably worse fit with $\chi^2=2.05$ where it fails to match the observed Na, Mg, Ti, Cr, and Mn abundance within $1\sigma$ uncertainty with deviations of $1.58\sigma$, $2.47\sigma$, $1.60\sigma$, $2.05\sigma$, and $1.27\sigma$, respectively.
The best-fit single CCSN model also provides a similar poor fit with $\chi^2=2.15$ where it fails to match Na, Mg, Ti, Cr, and Fe abundance within $1\sigma$ uncertainty with deviations of $2.31\sigma$, $1.72\sigma$, $2.04\sigma$, $2.28\sigma$, and $1.03\sigma$, respectively, where near-\ch SN 1a contributes significantly only to Mn. 
The best-fit CCSN+PISN model is effectively the same as the best-fit single CCSN model as there is no contribution from PISN. The best-fit single PISN model provides a very poor fit and can be essentially ruled out. 

Overall, the best-fit CCSN+sub-\ch model provides a good fit to the observed abundance pattern while the best-fit 2CCSNe model gives an acceptable fit. The best-fit single CCSN, CCSN+near-\ch, and CCSN+PISN models all provide poor fits. Because there is no contribution from PISN in the best fit CCSN+PISN model and the single PISN provides a very poor fit, it indicates a complete lack of PISN signature in the abundance pattern observed in Scl11\_1\_4296.

Although the best-fit CCSN+sub-\ch model provides a good quality fit and is the overall best fit, it is unclear whether an association with sub-\ch SN 1a can be established. First, as noted earlier, in the best-fit CCSN+sub-\ch model, the fact that sub-\ch SN 1a accounts for almost all of the Fe is inconsistent with the super solar $\B{Sc}{Fe}$. Second, the best-fit CCSN+sub-\ch model has a relatively large deviation of $2.3\sigma$ for Cr. Third, the overall quality of fit is not substantially better than the 2CCSNe model. Lastly, the super-solar $\B{C}{Fe}$ measured in this star (even after internal depletion) does not provide any additional support for SN 1a signature in sharp contrast to SDSSJ0018-0939.
For this reason, although there are indications of sub-\ch SN 1a contribution, the signature is unclear. In this regard, it is important to note that from the point of view of chemical evolution, due to the extremely low metallicity of $\B{Fe}{H}=-3.77$, the contribution from SN 1a is much less likely than pure CCSN scenarios.

\subsubsection{J1458+1128 (Fig.~\ref{fig:groupC_2}b)}
J1458+1128 is an EMP star, with metallicity $\B{Fe}{H}=-3.6$  with 8 elements detected with $Z\leq30$~\citep{Li2022ApJ}. It has unusually low Ca with $\B{Ca}{Fe}=-1.1$ which cannot be fit by any of the sources leading to an overall poor fit. 
%It also has super-solar values of $\B{C}{Fe}$ and $\B{Ti}{Fe}$.
The best-fit model from the CCSN+sub-\ch scenario provides the overall best fit with $\chi^2=1.91$ where it can match all the elements within $1\sigma$ uncertainty but has a large deviation for Ca of $3.33\sigma$ and a minor deviation for Ti of $1.22\sigma$. The best-fit model is a combination of ejecta from the \texttt{z10.6}-$Y_e$ model with some fallback and the sub-\ch SN 1a model M09\_05, where the latter contributes dominantly to most elements from Si--Mn. 
The best-fit 2CCSNe model provides a worse fit with $\chi^2=2.82$ where it also fails to match Ca and Ti but with even higher deviations of $3.87\sigma$ and $2.04\sigma$, respectively. The quality of fit from the best-fit single CCSN model is considerably worse with $\chi^2=3.27$ where it additionally fails to match Cr and Fe. The best-fit CCSN+near-\ch and CCSN+PISN models are effectively the same as the single CCSN as they have no contribution from the non-CCSN counterpart. The single PISN provides by far the worst fit with a $\chi^2=28.07$.

Overall, all scenarios provide poor fits except for the very poor fit from the single PISN scenario. The main reason is the highly sub-solar $\B{Ca}{Fe}=-1.1$. Given the importance of Ca abundance for this star, it should be re-evaluated to confirm whether such extremely low Ca abundance is indeed correct. Because of the poor quality of fit from all scenarios, no clear signature of any source is found.

\subsubsection{CC10690 (Fig.~\ref{fig:groupC_2}c)}
CC10690 has a metallicity of $\B{Fe}{H}=-1.96$ with 7 elements detected with $Z\leq30$~\citep{norris2017ApJS}.  Although the metallicity is marginally higher than the maximum metallicity of  $\B{Fe}{H}=-2$ for VMP stars, we included this star in our analysis similar to SMSSJ034249-284215.
The best-fit CCSN+sub-\ch model provides the overall best fit with $\chi^2=0.88$ which is a combination of ejecta from the \texttt{z13.7}-$S_4$ model with negligible fallback, and sub-\ch SN 1a model with low mass CO core model M08\_05. In this case, SN 1a contributes substantially to most of the elements from Sc--Mn. 
This model provides a very good fit and can match most elements within $1\sigma$ uncertainty with Ti, Cr, and Fe as minor outliers with deviations of $1.38\sigma$, $1.10\sigma$, and $1.10\sigma$, respectively. 
The best-fit 2CCSNe model can also provide almost an equally good fit with a $\chi^2=0.96$ with a slightly higher deviation of $1.67\sigma$ for Ti that is compensated by a lower deviation of $1.08$ for Fe. This model is a combination of ejecta from \texttt{z27}-$Y_e$ model without fallback and \texttt{z29}-$Y_e$ model with minimal fallback both resulting from a high energy explosion of $1.2\times 10^{52}\,\erg$.
The best-fit single CCSN model provides a slightly worse fit due to a somewhat higher Ti deviation of $2.12\sigma$ but provides a good fit overall. The best-fit CCSN+near-\ch and CCSN+PISN models have no contribution from non-CCSN sources and are effectively the same as the best-fit single CCSN model. The single PISN model provides a very poor fit with $\chi^2=19.17$.

Overall, the CCSN+sub-\ch and 2CCSNe scenarios provide very good fits to the observed abundance pattern while the single CCSN scenario provides a good fit. Our analysis indicates that this star is not compatible with contributions from near-\ch SN 1a or PISN. Because both  CCSN+sub-\ch and 2CCSNe scenarios provide comparable fits, no clear signature of either CCSN or near-\ch SN 1a can be claimed.

 \subsubsection{HE2148-2039 (Fig.~\ref{fig:groupC_2}d)}
HE2148-2039 is an EMP star of metallicity $\B{Fe}{H}=-3.30$ with only 6 elements detected with $Z\leq30$~\citep{purandardas2021ApJ}.
The best-fit CCSN+sub-\ch model provides a perfect fit with $\chi^2=0.0$, which is a combination of low mass CCSN model \texttt{10.9}-$Y_e$ with some fallback and sub-\ch SN 1a model M09\_10 where the latter contributes substantially to almost all elements from Si--Zn. The best-fit models from all other scenarios (except single PISN) also provide excellent fits for the abundance pattern observed in HE2148-2039. The best-fit models from CCSN+near-\ch and CCSN+PISN are effectively the same as the best-fit single CCSN model as there is no contribution from either near-\ch or PISN in the best-fit (see $\eta$ in Fig.~\ref{fig:groupC_2}d). 

Overall, all scenarios except the single PISN scenario provide very good fits to the observed abundance pattern. Because near-\ch SN 1a and PISN do not contribute to the best-fit CCSN+near-\ch and CCSN+PISN models, respectively, it indicates a lack of any near-\ch or PISN signatures in the limited set of elements observed in HE2148-2039. More elements are needed to be detected in order to find the most likely source of this star but the single PISN scenario is conclusively ruled out.  Because multiple scenarios provide very good fits, no clear signature of any particular source can be claimed.  

\begin{table*}
\centering
\caption{Summary of quality of fit from all six scenarios for each star: very good (VG), good (G), acceptable (A), poor (P), and very poor (VP). The scenario that provides the overall best fit for each star is highlighted with *. The level of contribution from SN 1a and PISN in CCSN+SN 1a and CCSN+PISN are also listed: high (H), medium (M), low (L), and negligible (N).}
    %\resizebox{\textwidth}{!}{%
    \begin{tabular}{|c| c |c |c |c |c |c |c|}
    \hline
         Star               &PISN     &CCSN     &2CCSNe     &CCSN+        &CCSN+         &CCSN+       & Signature \\
                            &         &         &           &near-\ch     &sub-\ch       &PISN        & \\
         \hline 
       HE0007-1752           &VP       &G        &*VG        &G,N          &G,M           &G,M         &hints of pure CCSN\\
       \hline
       SMSSJ034249-284215   &VP       &P        &*A         &A,M          &P,M           &P,L         &None\\
       \hline
       HE1207-3108          &VP       &G/A        &*VG        &VG,H         &G,L           &VG,L        &None\\
       \hline
       J0025+2305           &VP       &A        &*VG        &A,N          &VG,H          &G,M       &None\\
       \hline
       J09084+3119          &VP       &P        &*A         &P,N          &P,L           &P,N         &hints of pure CCSN\\
       \hline
       J1151-0054           &VP       &VG       &*VG        &VG,N         &VG,L          &VG,N        &hints of pure CCSN\\
       \hline
       SDSSJ0254+3328       &VP       &*VG      &*VG        &*VG,M        &*VG,H         &*VG,H    &None\\
       \hline
       SDSSJ1633+3907       &VP       &VG       &*VG        &*VG,M        &*VG,H         &*VG,M     &None\\
       \hline
       HE0533-5340          &VP       &P        &P          &*P,H         &P,H           &P,H        &None\\
       \hline
       J1542+2115           &VP       &G/A      &G/A        &*G,H       &G/A,L         &G/A,N        &hints of near-\ch SN 1a\\
       \hline
       SDSSJ0018-0939       &VP       &P        &P          &P,H          &*VG,H         &P,N          &clear sub-\ch SN 1a signature\\
       \hline
       J1010+2358 \citepalias{skuladottir2024ApJ}     &VP       &VG       &VG         &VG,L         &*VG,H         &VG,N         &None\\
       \hline
       J1010+2358 \citepalias{Thibodeaux2024}     &VP       &VG       &VG         &VG,M         &*VG,H         &VG,N         &None\\
       \hline
       ET0381               &VP       &G/A      &G/A        &G/A,H        &*VG,H         &G/A,N        & sub-\ch SN 1a signature\\
       \hline
       SCl11-1-4296         &VP       &P        &A          &P,M          &*G,H          &P,N          &hints of sub-\ch SN 1a\\
       \hline
       J1458+1128           &VP       &P        &P          &P,N          &*P,H          &P,N          &None\\
       \hline
       CC10690              &VP       &G        &VG         &G,N          &*VG,M         &G,N          &None\\
       \hline
       HE2148-2039          &VP       &VG       &VG         &VG,N         &*VG,H         &VG,N         &None\\
       \hline     

    \end{tabular}%}
   
    \label{tab:alpha_poor_summary}
\end{table*}

\section{Summary and Conclusions}
We analyzed 17 $\alpha$PVMP  stars by matching the observed abundance pattern with the yields from 6 different theoretical scenarios that cover the range of possible sources in the early Galaxy that include PISN, CCSN, and SN 1a. Table~\ref{tab:alpha_poor_summary} gives the overall summary of the best-fit analysis where the quality of fit from all 6 scenarios for each star is listed. The quality of fit is classified as very good (VG), good (G), acceptable (A), poor (P), and very poor (VP) as discussed in the detailed analysis presented above for each star. Additionally, we also quantify the contribution of near-\ch SN 1a, sub-\ch SN 1a, and PISN in the best-fit models from CCSN+near-\ch, CCSN+sub-\ch, and CCSN+PISN, respectively, as high (H), medium (M), low (L), and negligible (N). Among all the 17 stars, the best-fit single PISN models uniformly provide a very poor fit and can be strongly ruled out. Among the 5 remaining scenarios, we effectively have 4 scenarios as the single CCSN models are already included as part of the other scenarios corresponding to $\alpha\approx 0$. Out of the remaining 4 scenarios, at least one of them can provide a fit whose quality is acceptable or better, i.e., acceptable, good or very good, for all stars except HE0533-5340 and J1458+1128. For these two stars, all 4 scenarios provide a poor fit which is primarily due to extremely peculiar abundance(s) of one or two elements. For HE0533-5340, the reason for poor fit is the super-solar $\B{Mn}{Fe}$ along with highly sub-solar $\B{Cr}{Fe}$, whereas for J1458+1128 the highly sub-solar $\B{Ca}{Fe}$ leads to a poor fit. 

Concerning the likely origin of the remaining 15 stars, we consider the results from the best-fit models from the 2CCSNe scenario and the three other scenarios that combine ejecta from CCSN and a non-CCSN source i.e., CCSN+near-\ch, CCSN+sub-\ch, and CCSN+PISN.
We find that the best-fit 2CCSNe model can match the abundance pattern of 14 stars ($82\,\%$) with a fit whose quality is acceptable or better (4 acceptable, 1 good, 9 very good), out of which it is the overall best fit for 8 stars. However, a clear signature of CCSN is difficult to establish since for almost all stars at least one of the other three scenarios can also provide comparable fits with non-negligible contribution from the non-CCSN source. 
The exceptions to this are J09084+3119 and J1151-0054.  For J09084+3119, only 2CCSNe provides an acceptable fit whereas the three other scenarios not only provide poor fits but also the contribution from non-CCSN source is either negligible (near-\ch SN 1a and PISN) or low (sub-\ch SN 1a). For J1151-0054, although the other three scenarios provide good fits, the contribution from non-CCSN sources is low. Overall, for these two stars, although there are indications of a pure CCSN origin, no clear signature can be claimed particularly due to the very few elements detected in these two stars. Among the Group A stars, HE0007-1752 is the only star that has multiple elements detected where the 2CCSNe provides a very good fit compared to other scenarios which provide good fits. While this hints towards an origin from gas polluted by pure CCSN ejecta, it is far from a clear signature. 

With regard to the other three scenarios, we only consider the cases where the non-CCSN source has a non-negligible contribution to the best-fit model. We find that the best-fit CCSN+sub-\ch models can match the abundance pattern of 13 stars ($76\,\%$) with a quality of fit that is acceptable or better (1 acceptable, 3 good, 9 very good) out of which it is the overall best-fit for 7 stars.
Out of these 13 stars, the contribution of sub-\ch SN 1a is substantial for 10 stars and low for the remaining 3 stars. We find that SDSSJ0018-0939 has the clearest signature of a sub-\ch SN 1a among all the 13 stars where only the best-fit CCSN+sub-\ch model can provide a very good fit whereas the other three scenarios provide poor fits. The situation is somewhat similar for ET0381 where only the best-fit CCSN+sub-\ch model can provide a very good fit. 
However, in this case, the other scenarios can provide a quality of fit that can be classified as somewhere between acceptable and good. Additionally, the abundance of the Ti, which is a key element responsible for the association of sub-\ch SN 1a, is uncertain and the SN 1a signature of highly sub-solar $\B{C}{Fe}$ present in SDSSJ0018-0939 is not found in this star. For these reasons, although ET0381 has a reasonably clear signature of sub-\ch SN 1a, it cannot be considered as a near-smoking-gun signature in contrast to SDSSJ0018-0939. We find that Scl11-1-4296 also has some hints of sub-\ch SN 1a where only the best-fit CCSN+sub-\ch model can provide a good fit. However, the fact that it fails to match Cr with a somewhat large deviation of $2.35\sigma$, and the fact that the best-fit 2CCSNe model can provide an acceptable fit that is not much worse, makes the signature of sub-\ch SN 1a unclear.

The best-fit CCSN+near-\ch model can match the abundance pattern of 7 stars ($41\,\%$) with a quality of fit that is acceptable or better (2 acceptable, 1 good, and 4 very good) where near-\ch SN 1a contribution is substantial for all 7 stars.
Of the 7 stars,  CCSN+near-\ch provides the overall best fit for only 1 star, namely J1542+2115. Although this star has some hints of near-\ch SN 1a, it does not have a clear signature as the other three scenarios can provide acceptable fits. 
The best-fit CCSN+PISN models can match the abundance for 5 stars ($29\,\%$) with a quality of fit that is acceptable or better (2 good, 3 very good) with a substantial contribution from PISN in 4 stars and a minor contribution in 1 star. However, this scenario does not provide the overall best-fit for any of the stars highlighting the lack of PISN signature in $\alpha$PVMP stars. 

The results presented in this work show that $\alpha$PVMP stars are not exclusively associated with SN 1a. On the contrary, we find that pure CCSN ejecta  (2CCSNe scenario) can account for the highest fraction of $\alpha$PVMP stars (14 out of 17 stars) which is followed closely by the CCSN+sub-\ch scenario (13 out of 17 stars). This shows that both pure CCSN ejecta or a combination of CCSN and sub-\ch SN 1a ejecta can explain the abundance pattern observed in $\alpha$PVMP stars.
%these two are the most likely scenarios for explaining the abundances in $\alpha$PVMP stars.  
Some of the $\alpha$PVMP stars could also have been formed from gas polluted by CCSN and near-\ch SN 1a which is evident from the fact that CCSN+near-\ch scenario can provide an acceptable (or better) fit to 7 of the 17 stars. Although pure PISN ejecta can be strongly ruled out, the possibility of some PISN contribution cannot be ruled out for some of the $\alpha$PVMP stars as the CCSN+PISN scenario can fit 5 of the 17 stars.

Interestingly, if we take the relative fraction of $\alpha$PVMP stars that can be explained by sub-\ch and near-\ch SN 1a as an indicator of their relative frequency in the early Galaxy, it would imply that sub-\ch SN 1a is roughly twice as frequent in the early Galaxy than near-\ch SN 1a. This would be consistent with the findings from Galactic chemical evolution studies of Mn and Ni by ~\citet{eitner2020,eitner2023} where they found that $\sim 75\%$ of all SN 1a in the early Galaxy need to be from sub-\ch SN 1a to be consistent with the observations. This would also be consistent with the results from ~\citet{kirby2019ApJ,reyes2020ApJ} where they find that sub-\ch SN 1a is the dominant SN 1a channel in early dwarf galaxies.    

Although, for each star, one would ideally like to find a clear signature of a specific source, it is extremely difficult to establish such a signature since more than one source can provide comparable fits for a star. Our analysis did not find a clear signature for CCSN, PISN, and near-\ch SN 1a in any of the 17 stars. For sub-\ch SN 1a, only one star showed a clear unambiguous signature while another showed a reasonably clear signature.
In general, the detection of more elements could help break this degeneracy, particularly in stars with few detections such as SDSSJ0254+3328 and SDSSJ1633+3907. However, the detection of multiple elements is not guaranteed to result in a clear identification of a source as is evident from stars such as J1010+2358 and HE1207-3108. 

The primary reason for a lack of clear signature of either pure CCSN or another non-CCSN source is related to the abundance patterns from the ejecta from these sources. 
For PISN resulting from massive progenitors with He core mass of $\gtrsim 100\,\Msun$, the value of $\B{X}{Fe}$ for even $Z$ and Fe peak elements is not very different from CCSN with low-fallback. The extremely low abundance of odd $Z$ elements in such progenitors is what makes it distinct from CCSN and is the primary reason why we can definitively exclude the single PISN scenario for most stars as they do not show such a feature. 
For lighter mass PISN progenitors, there is a large enhancement of light and intermediate $\alpha$ elements relative to Fe which is also distinct from CCSN but incompatible with $\alpha$-poor pattern which again leads to the exclusion of the single PISN scenario. Thus, in order to get a good fit from the CCSN+PISN scenario, a considerable contribution of CCSN is required which dilutes the clear PISN signatures and makes the final abundance pattern degenerate with CCSN patterns. 

The situation is somewhat similar for the near-\ch SN 1a. Because these models cannot produce elements below Si, a large CCSN contribution is required in the CCSN+near-\ch models to explain elements such as Mg and Na. When the CCSN ejecta, which does not undergo large fallback, is mixed with near-\ch SN 1a ejecta, the only unique feature that survives is the solar $\B{Mn}{Fe}$ along with sub-solar $\B{X}{Fe}$ values of elements lighter than Si. However, other than solar $\B{Mn}{Fe}$, in most cases, pure CCSN ejecta can also produce similar sub-solar $\B{X}{Fe}$ making the patterns from the two scenarios roughly degenerate. It is important to note that when ejecta from CCSN, which undergoes a large fallback of the innermost regions containing all the Fe peak elements, is mixed with near-\ch SN 1a ejecta, the latter will dominate the Fe peak. Although such models are already included in the CCSN+near-\ch scenario, it does not show up as a good fit for any of the stars. 

The same arguments also apply to sub-\ch SN 1a but the prospects of finding a clear signature are much better. This is because sub-\ch SN 1a resulting from the explosion of CO core mass of $\lesssim 1 \Msun$, have super-solar $\B{X}{Fe}$ for Ti--Cr (due to He shell detonation) while crucially still satisfying $\B{Mg}{Fe}<0$. Such a signature cannot be replicated by pure CCSN or PISN ejecta. Importantly, even with substantial CCSN contribution in the final mixed ejecta, this signature survives and can be used to find a clear sub-\ch SN 1a signature as in the case of SDSSJ0018-0939. 

We also note,  that in addition to SN 1a features such as solar $\B{Mn}{Fe}$ or super-solar $\B{X}{Fe}$ for Ti--Cr, if a $\alpha$PVMP star has highly sub-solar values of $\B{X}{Fe}\lesssim -0.6$ for C (that is not due to internal depletion) and O, it can be a strong indicator of a major contribution from either near-\ch or sub-\ch SN 1a as such low $\B{C}{Fe}$ and $\B{O}{Fe}$ values is not possible from pure CCSN ejecta. In such cases, the quality of fit from the best-fit CCSN+SN 1a model will be substantially better than the best-fit from other scenarios and a clear signature of SN 1a can be claimed as in the case of SDSSJ0018-0939. In this regard, it is important to note $\B{X}{Fe}\lesssim -0.6$ for C and O is also possible from high mass PISN progenitors but the pattern near the Fe peak is distinct from SN 1a signatures.

Lastly, we note that the number of $\alpha$PVMP stars is currently very few. With an increasing number of detections of VMP stars, the number of $\alpha$PVMP stars will increase. Detailed abundance patterns in such stars will be an important way to look for a clear signature of SN 1a and particularly sub-\ch SN 1a. Such stars in turn will provide direct constraints on the nucleosynthesis that could provide crucial insights into the nature of the SN 1a explosion mechanism and the relative frequency of near-\ch and sub-\ch SN 1a. In this regard, isotopic abundances of elements could be crucial in breaking the degeneracy in the elemental abundances from different scenarios to facilitate the detection of a clear signature of a particular source. Although measurement of isotopic abundances in VMP stars is difficult, for some of the elements, it could be within reach of upcoming large telescopes such as the Extremely Large Telescope, the Giant Magellan Telescope, and the Thirty Meter Telescope. We plan to explore this in future. 

\section*{Data Availability}
Data is available upon reasonable request.

\bibliography{main}
\bibliographystyle{aasjournal}
\end{document}